\documentclass[journal,onecolumn,twoside,draftcls,12pt]{IEEEtran}

\usepackage{cite}
\usepackage{citesort}
\usepackage{graphicx,amssymb}
\usepackage{textcomp,gensymb}
\usepackage{amsmath}
\usepackage{bbding}
\usepackage{multirow}
\usepackage{multicol}
\usepackage{latexsym}
\usepackage{url}
\usepackage[usenames,dvipsnames,svgnames,table]{xcolor}
\usepackage{rotating}
\usepackage{longtable}
\usepackage{longtable,lscape}
\usepackage{url}
\usepackage{array}
\usepackage{afterpage}
\usepackage{pdflscape}
\usepackage{lipsum}
\usepackage{gensymb}
\usepackage{bbding} 
\usepackage{float}
\usepackage{stfloats}
\usepackage{balance}
\usepackage{epstopdf} 
\usepackage[ulem=normalem]{changes}
\usepackage{caption}
\usepackage[labelformat=simple]{subcaption}

\usepackage[nonumberlist,acronym,shortcuts]{glossaries}
\makeindex
\makeglossaries
\newcolumntype{L}[1]{>{\raggedright\let\newline\\\arraybackslash\hspace{0pt}}m{#1}}
\newcolumntype{C}[1]{>{\centering\let\newline\\\arraybackslash\hspace{0pt}}m{#1}}
\newcolumntype{R}[1]{>{\raggedleft\let\newline\\\arraybackslash\hspace{0pt}}m{#1}}
\usepackage{tabularx}
\usepackage[outerbars,xcolor]{changebar}
\ifCLASSINFOpdf

\else

\fi
\begin{document}
\title{Differential Evolution Algorithm Aided Turbo Channel Estimation and Multi-User
 Detection for G.Fast Systems in the Presence of FEXT}
\author{Jiankang~Zhang,~\IEEEmembership{Senior Member,~IEEE},
        Sheng~Chen,~\IEEEmembership{Fellow,~IEEE},
        Rong~Zhang,~\IEEEmembership{Senior Member,~IEEE}, \\
        Anas F. Al Rawi,~\IEEEmembership{Member,~IEEE},
        Lajos~Hanzo,~\IEEEmembership{Fellow,~IEEE}
\thanks{J.~Zhang, S.~Chen, R.~Zhang and L.~Hanzo are with School of Electronics and
 Computer Science, University of Southampton, Southampton SO17 1BJ, U.K. (Emails:
 jz09v@ecs.soton.ac.uk, sqc@ecs.soton.ac.uk, rz@ecs.soton.ac.uk, lh@ecs.soton.ac.uk).
 S.~Chen is also with King Abdulaziz University, Jeddah 21589, Saudi Arabia.} %
\thanks{A.~F.~Al Rawi is with British Telecommunications, Martlesham, Ipswich, U.K.
 (Email: anas.mohsin@bt.com).} %
\thanks{The financial support of the European Research Council's Advanced Fellow Grant, 
 the Royal Society Wolfson Research Merit Award  as well as of the EPSRC project
 EP/N004558/1 are gratefully acknowledged. The research
data for this paper is available at https://doi.org/10.5258/SOTON/D0550.}
\vspace{-5mm}
}
\maketitle

\IEEEpeerreviewmaketitle

\begin{abstract}
 The ever-increasing demand for broadband Internet access has motivated the further
 development of the  digital subscriber line to the G.fast standard in order to expand
 its operational band from 106\,MHz to 212\,MHz. Conventional far-end crosstalk (FEXT)
 based cancellers falter in the upstream transmission of this emerging G.fast system.
 In this paper, we propose a novel differential evolution algorithm (DEA) aided turbo
 channel estimation (CE) and multi-user detection (MUD) scheme for the G.fast upstream
 including the frequency band up to 212\,MHz, which is capable of approaching the optimal
 Cramer-Rao lower bound of the channel estimate, whilst approaching the optimal maximum
 likelihood (ML) MUD's performance associated with perfect channel state information,
 and  yet only imposing about 5\% of its computational complexity. Explicitly, the turbo
 concept is exploited by iteratively exchanging information between the continuous
 value-based DEA assisted channel estimator and the discrete value-based DEA MUD. Our
 extensive simulations show that 18\,dB normalized mean square error gain is attained by
 the channel estimator and 10\,dB signal-to-noise ratio gain  can be achieved by the MUD
 upon  exploiting this iteration gain. We also quantify the influence of the CE error,
 of the copper length and of the impulse noise. Our study demonstrates that the proposed
 DEA aided turbo CE and MUD scheme is capable of offering near-capacity performance at an
 affordable complexity for the emerging G.fast systems. 
\end{abstract}


\section{Introduction}\label{S1}

 The demand for high-speed broadband Internet access has motivated the construction of
 the hybrid digital subscriber line (DSL) and optical fiber infrastructure, which has
 been standardized as G.fast by the International Telecommunication Union Telecommunication
 Standardization Sector (ITU-T) \cite{oksman2016itu}. G.fast still relies on the copper
 access network and in-premises wiring for its last hundred meters, because it is
 prohibitively expensive to replace copper by optical fiber. The G.fast standard has
 already exploited a broad spectrum spanning up to 106\,MHz, while the band up to 212\,MHz
 has been planed for future broadband access \cite{effenberger2016future,odling2009fourth}.
 However, exploiting the spectrum beyond 30 MHz inevitably imposes significant
 electromagnetic coupling between the neighboring twisted-pairs, which is referred to as
 crosstalk. There are two types of crosstalk, depending on the specific source of coupling.
 Explicitly, coupling originating from transmitter co-located with a receiver is called
 near-end crosstalk (NEXT), while the coupling arriving from  the opposite side of the
 duplex link is known as far-end crosstalk (FEXT) \cite{binyamini2013adaptive}. NEXT can
 be avoided by employing frequency duplexing division and transmission synchronization
 for separating the downstream and upstream transmissions \cite{binyamini2013adaptive},
 or by vectoring the source signals for the sake of increasing the total throughput of
 the cable \cite{ginis2002vectored}. By contrast, FEXT remains a significant impairment
 that hampers achieving high data rates for G.fast systems \cite{oksman2016itu}.

\begin{figure*}[bp!]
\vspace{-3mm}
\begin{center}
\includegraphics[width=0.45\textwidth,angle=0]{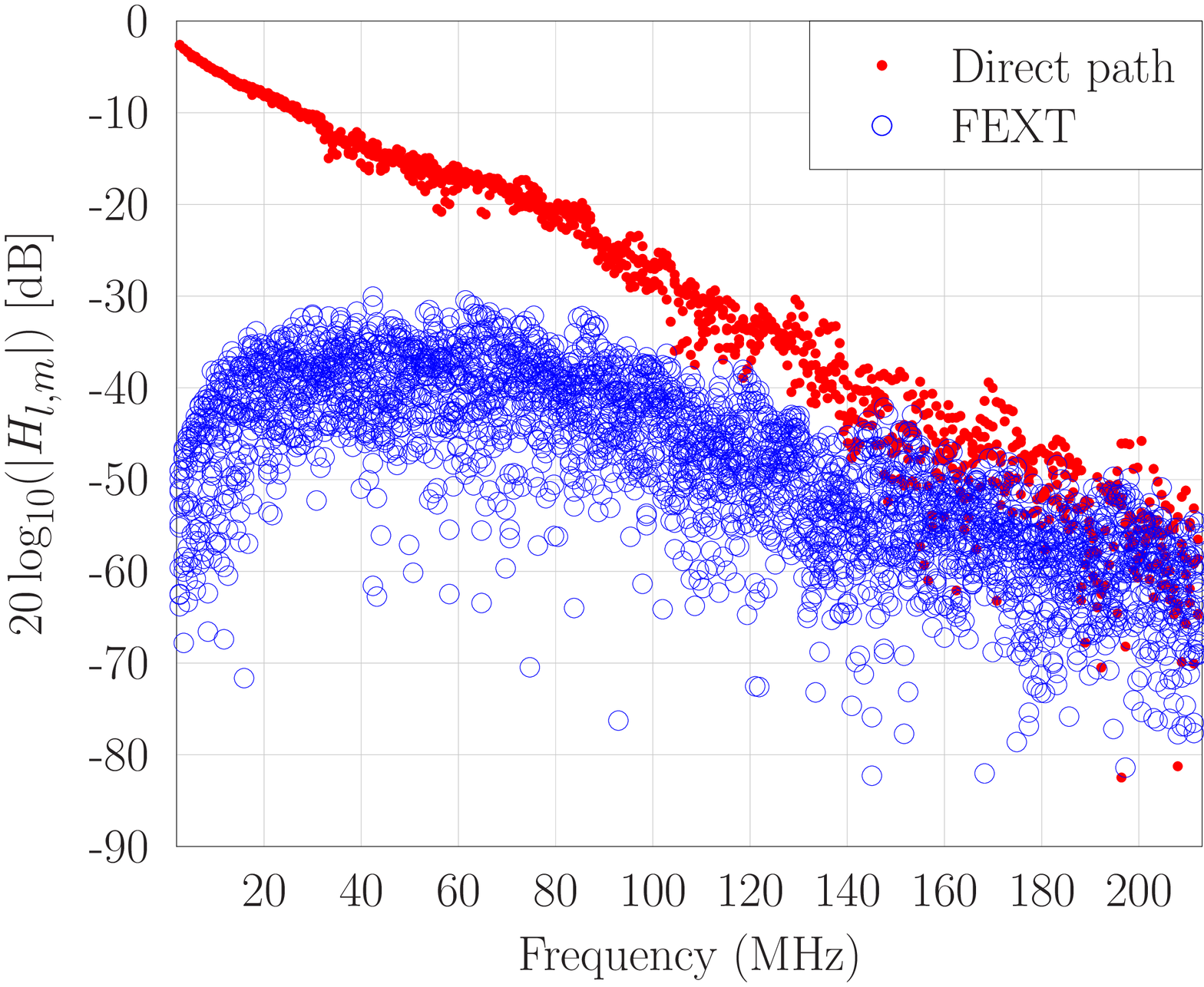}\hspace*{6mm}\includegraphics[width=0.45\textwidth,angle=0]{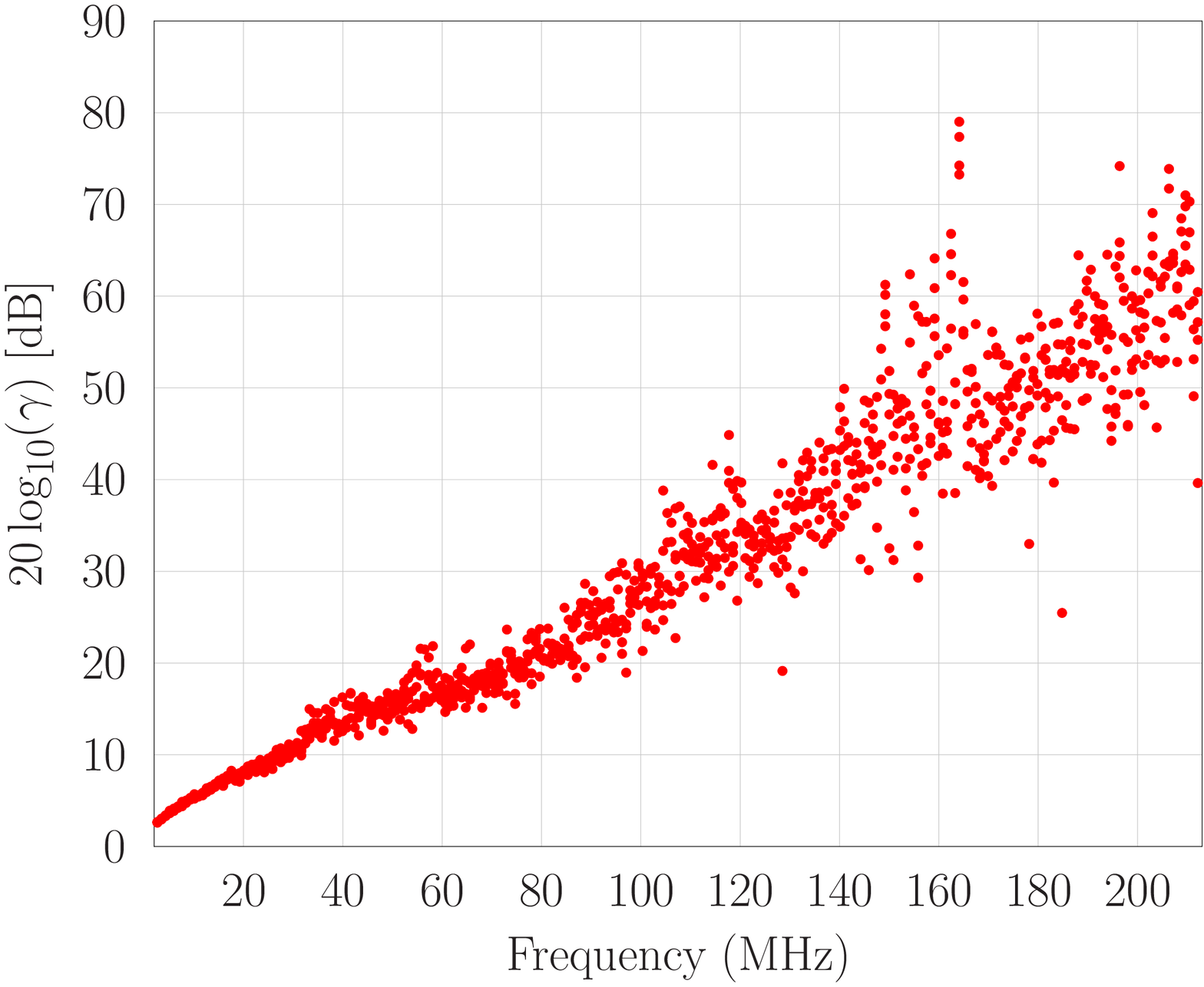}
\end{center}
\vspace{-5mm}
\begin{center}\hspace*{7mm}{\small (a)}\hspace*{80mm}{\small (b)}\end{center}
\vspace{-4mm}
\caption{\small (a)~Measured direct channel and crosstalk channel strengths, and (b)~the
 average noise power amplified by the ZF FEXT canceller.}
\label{FIG1}
\vspace{-1mm}
\end{figure*}

\subsection{Existing solutions for mitigating FEXT in the downstream}

 FEXT is generally mitigated by using spectrum shaping and vectoring \cite{zanko2016gigabit}
 in the downstream transmission. The vectoring technique has also been standardized in
 G.993.5 by the ITU-T \cite{ITU-T2010Self} and it was further developed with the objective of
 approaching Gigabit rates \cite{oksman2016itu}. More specifically, in the downstream, the
 transmitters are co-located at the central office (CO) or at the optical network unit (ONU).
 Hence vectoring (precoding) can be applied to the transmitted signals. As a powerful
 nonlinear precoding scheme, the Tomlinson-Harashima precoder (THP) \cite{ginis2000multi} was
 demonstrated to approach the single-user bound for a  bandwidth of up to 17.6\,MHz. However,
 the THP is incompatible with the previous version of G.fast. Hekrdla \emph{et al.}
 \cite{hekrdla2015ordered} developed a dynamic ordering-based THP by taking into account the
 specific G.fast channel statistics. Zhang \emph{et al.} \cite{zhang2017expanded} conceived
 the concept of expanded constellation mapping for maximizing the received signal power, while
 cancelling the FEXT. However, the THP imposes a high computational complexity both on the
 transmitter and receiver. The investigations of \cite{Maes_etal2016} revealed that the THP
 was also sensitive to the channel state information (CSI)  estimation error. By comparison,
 low-complexity linear precoding used in the context of very-high-bit-rate DSL (VDSL) exhibits
 almost the same performance as the more complex nonlinear ones \cite{cendrillon2006near}.
 The vectoring concept of \cite{ginis2002vectored} was proposed for cancelling the FEXT by
 exploiting user coordination at the either CO or ONU. Furthermore, for the second-generation
 VDSL (VDSL2) system \cite{eriksson2006vdsl2}  zero-forcing (ZF) precoding was adopted to
 mitigate the FEXT based on the column-wise diagonal dominant (CWDD) nature of the copper
 channel \cite{ginis2002vectored,zafaruddin2015performance}, which was demonstrated to be
 near-optimal below 17.6\,MHz. However, ZF precoding relies on inverting a matrix at each
 tone, which imposes an excessive computational complexity, particularly for a large number
 of copper pairs in a cable. Leshem \emph{et al.} \cite{leshem2007low} simplified the ZF
 precoder by only exploiting the first-order and second-order statistics of the CSI. As a
 further development, Zanko \emph{et al.} \cite{zanko2016gigabit} simplified the ZF precoder
 by adopting the least mean square algorithm. Adaptive precoders
 \cite{louveaux2006error,louveaux2006adaptive,louveaux2010adaptive} were also designed for
 cancelling the FEXT by only exploiting the polarity of the symbol errors observed. Moreover,
 a potent combination of precoding and dynamic spectrum management was also conceived for
 downstream vectored transmission. Specifically, Lanneer  \emph{et al.} \cite{lanneer2017linear}
 developed both a linear and a nonlinear precoding-based dynamic spectrum management, which
 maximizes the weighted sum-rate under realistic per-line total power and per-tone spectral
 mask constraints. Note that  the FEXT-contaminated DSL system can be viewed as a multi-user
 multiple-input multiple-output (MIMO) system \cite{fazlollahi2016fext}. Therefore, the
 powerful multi-user transmission techniques originally developed for wireless MIMO
 communications, such as the vectors perturbation technique \cite{VPTx1,VPTx2}, can be
 readily be invoked for downstream transmission in G.fast systems.

\subsection{Existing solutions for mitigating FEXT in the upstream}

 By contrast, in the upstream, the transmitters are those of independent users at different
 locations. Since there are no physical lines connecting the distributed users, no
 coordination is possible amongst their distributed transmitters, and it is impossible to
 apply centralized transmit precoding techniques. Therefore, the FEXT is typically mitigated
 either at the CO or at the ONU by exploiting sophisticated FEXT cancellation techniques
 \cite{cheong2002multiuser,ginis2002vectored}. Explicitly, Im \emph{et al.} \cite{im2002fext}
 proposed a joint FEXT canceller and equalizer, while Hormis \emph{et al.}
 \cite{hormis2005monte} viewed the FEXT-impaired channel as a MIMO channel and proposed a
 soft interference canceller for quad-wire loops based on the sequential Monte Carlo
 technique. Ginis \emph{et al.} \cite{ginis2002vectored} proposed to successively decode the
 received signal based on the QR decomposed channels and on the previous decision, which may
 be viewed as a special case of the ZF aided  generalized decision feedback equalizer
 (ZF-GDFE) \cite{cioffi1997generalized}. The achievable performance of the ZF-GDFE critically
 depends on the decoding order used \cite{chen2007optimized}, and to achieve its full
 performance potential, exhaustive search is required, which may impose an excessive
 complexity. Chen \emph{et al.}  \cite{chen2007optimized} considerably reduced the
 computational complexity either with the aid of an efficient successive ordering search or
 by a  modified greedy search. Their results show that the successive ordering search aided
 ZF-GDFE is  capable of approaching the rate of the optimal ordering based ZF-GDFE. For
 VDSL2 systems, the linear ZF equalizer of \cite{cendrillon2006near} is capable of closely
 approximating the performance of the ZF-GDFE, despite its lower complexity. The family of
 ZF-type FEXT cancellers treats the alien cross-talk as one of the self-FEXT contributions,
 which results in poor performance in the presence of alien noise. Zafaruddin  \emph{et al.}
 \cite{zafaruddin2015performance} proposed a constrained linear minimum-output energy
 receiver for cancelling both the self-crosstalk and the alien crosstalk in the VDSL
 upstream. Biyani \emph{et al.} \cite{biyani2013co} proposed to whiten the alien noise
 contaminating the VDSL systems with the aid of  a co-operative alien noise cancellation
 algorithm. Explicitly, the co-operative alien noise cancellation algorithm succeeds in
 removing the alien noise that persists after the ZF-FEXT canceller by invoking a
 sophisticated recursive scheme, which is capable of meeting the Cramer-Rao lower bound
 (CRLB), provided that the symbol decision errors are perfectly known, but naturally its
 performance will erode in the face of imperfect symbol decision error knowledge.

 Similar to the ZF precoder of the downstream, the ZF-based FEXT canceller of the upstream
 also requires matrix inversion at each tone of the multi-user DSL systems for  attaining
 a near-optimal performance. However, unlike the ZF precoder, a ZF-FEXT canceller, which
 basically relies on a ZF detection algorithm, will significantly enhance the additive noise
 power. Quantitatively, Fig.~\ref{FIG1}\,(b) depicts the average noise power amplified by
 the ZF-FEXT canceller, where we have $\gamma = \frac{\tilde{\sigma}^{2}}{\sigma_{o}^{2}}$,
 with $\sigma_{o}^{2}$ being the power of the original noise term, while $\tilde{\sigma}^{2}$
 is the power of the noise output by the ZF-FEXT canceller.  Observe that the noise power
 is amplified quite dramatically at higher operational frequencies. At the time of writing,
 the ZF-FEXT canceller is successfully deployed in  G.fast systems operating in a bandwidth
 spanning up to 100\,MHz. However, in the near future, the bandwidth will be increased up
 to around 200\,MHz. Observe from Fig.~\ref{FIG1}\,(b)  that the noise enhancement in an
 operating bandwidth of 200\,MHz is 30\,dB higher than that at 100\,MHz. Clearly, the
 existing ZF-FEXT canceller fails to perform well in these future high-bandwidth G.fast
 systems because of this dramatically increased noise enhancement. Moreover, the direct
 channels are overwhelmed  by the crosstalk  in the G.fast systems in the frequency range
 spanning from 100\,MHz to 212\,MHz, as shown in Fig.~\ref{FIG1}\,(a). Hence the ZF-FEXT
 canceller is no longer near-optimal, since the channel matrix does not hold the property
 of being column-wise diagonal dominant. Hence, more powerful solutions for mitigating FEXT resisted in upstream G.fast systems are needed.

\subsection{Motivations and contributions}

The G.fast upstream system is reminiscent of a multi-user MIMO system, intuitively,
 the maximum likelihood (ML) multi-user detector (MUD) is expected to provide the
 ultimate optimal solution, albeit its computational complexity increases exponentially
 both with the number of users and with the modulation order. However, the ML-MUD is
 impractical for the G.fast system, which may support up to 24 users with the aid of a
 4096-quadrature amplitude modulation (QAM) constellation, because it would require
 $4096^{24}$ cost function evaluations \cite{oksman2016itu}.

 For wireless systems, evolutionary algorithms (EAs) have been extensively applied both
 for downlink precoder designs \cite{EATx1,EATx2,EATx3} as well as for uplink MUD designs
 \cite{EARx1,EARx2,EARx3,EARx4,EARx5,EARx6}. In particular, it has been demonstrated that
 the EA-aided MUD solutions are capable of approaching the optimal ML-MUD performance at 
 a fraction of the computational complexity imposed by the ML-MUD 
 \cite{EARx1,EARx2,EARx3,EARx4,EARx5,EARx6,EARx7}. Among the various EAs, the differential
 evolution algorithms (DEAs) \cite{price2005differential,qin2009differential} have been
 shown to be particularly powerful in joint iterative channel estimation (CE) and MUD.
 Explicitly, they are capable of approaching the CRLB of CE and the optimal ML-MUD
 performance associated with the perfect CSI at a fraction of the ML-MUD complexity
 \cite{EARx5,EARx6,EARx7}. Furthermore, turbo CE and MUD/decoder techniques 
 \cite{Turbo1,Turbo2,Turbo3,Turbo4,Turbo5,Turbo6,Turbo7,Turbo8} have been widely
 developed for powerful wireless systems, which are capable of achieving near-capacity
 performance at an affordable complexity. {\bf{The conceptual similarity between the G.fast 
 and the wireless multi-user uplink arises from the fact that the joint optimization of
 CE and MUD in the both systems relies on a similar multi-objective, multidimensional joint
 optimization problem associated  with continuous CE parameters and discrete MUD parameters.
 Hence, it is beneficial to appropriately adapt these reduced-complexity state-of-the-art
 wireless techniques to the G.fast systems, which is capable of jointly  detecting multi-user upstream signals relying on powerful central signal processing unit at the central office.}}

 Against this background, our novel contributions are:
\begin{itemize}
\item [1)]  We conceive a new turbo CE and
 MUD scheme relying on the DEA for the emerging family of G.fast systems having an 
 operational bandwidth spanning up to 212\,MHz, which is capable of approaching the CRLB
 of channel estimation as well as the optimal ML-MUD's performance associated with the
 perfect CSI, while only imposing a fraction of the ML-MUD complexity. 

\item [2)] The joint optimization problem of 
 turbo CE and MUD  proposed for the G.fast upstream is converted to iteratively procedure of a continuous-parameter
 DEA assisted channel estimator, which searches through the channel space to find the
 optimal CE solution, and a discrete-parameter DEA assisted MUD, which is capable of
 finding the optimal ML solution of the transmitted data. 

\item [3)]  Furthermore, the continuous DEA
 assisted channel estimator and discrete DEA aided MUD  iteratively exchange their
 extrinsic information to attain turbo gains for both the CE and MUD. Specifically, the
 reliably detected symbols are iteratively fed back to the channel estimator to be
 exploited together with the pilot symbols for further improving the accuracy of the
 estimated CSI, while the enhanced CSI estimates further improve the MUD's detection
 reliability. 

\item [4)]  We carry our extensive investigations for the convergence of DEA aided CE and DEA aided MUD, 
 the impacts of system bandwidth, DSL loop length, impulse noise and channel estimation error as well as the analysis of computational complexity. 
Our extensive investigations show that 18\,dB normalized mean square error
 (NMSE) channel estimation gain and 10\,dB MUD signal-to-noise ratio (SNR) gain can be
 achieved by exploiting this iteration gain. Furthermore, we quantitatively investigate
 the influence of both the CE error and of the copper length as well as the impact of
 impulse noise. Our results  confirm that the proposed DEA aided turbo CE and MUD scheme
 offers near-capacity performance at an affordable complexity for the emerging family of
 G.fast systems. 

\end{itemize}
 
 The rest of this paper is organized as follows. The upstream G.fast system model is 
 described in Section~\ref{S2}. Section~\ref{S3} is devoted to our DEA assisted turbo CE
 and MUD. The  CRLB of the channel estimate is derived in Section~~\ref{S4}. Our simulation results
 and discussions are presented in Section~\ref{S5}, whilst our concluding remarks are
 offered in Section~\ref{S6}.

\section{Upstream System Model}\label{S2}

 We consider the upstream of DSL within a bandwidth spanning up to 212\,MHz, which
 supports $L$ users simultaneously transmitting their signals to the CO. Discrete
 multi-tone modulation is employed by each user, which occupies $\Omega$ MHz bandwidth of
 $N_c$ subcarriers, each allocated $\Omega/N_{c}$ MHz. We assume that the cyclic prefix
 is sufficiently long and that the users synchronously transmit their signals. Thus, there
 is no intersymbol interference and no inter-carrier interference. Hence we can process
 the signals on a per tone basis. By omitting the tone index, the $L$-user received signal
 vector on the tone of interest can be written as \cite{ginis2002vectored}
\begin{align}\label{EQ1_re_sig} % eq1
 \boldsymbol{Y} =& \boldsymbol{H} \boldsymbol{X} + \boldsymbol{W} ,
\end{align}
 where $\boldsymbol{Y}\in \mathbb{C}^L$ is the received signal vector, $\boldsymbol{X}\in
 \mathbb{C}^L$ is the transmitted signal vector of the $L$ users, and $\boldsymbol{W}\in
 \mathbb{C}^L$ is the zero-mean white Gaussian noise vector with the covariance matrix
 $\sigma_{w}^2\boldsymbol{I}_L$ in which $\boldsymbol{I}_L$ is the $L\times L$ identity
 matrix, while $\boldsymbol{H}\in \mathbb{C}^{L \times L}$ is the frequency-domain channel
 matrix whose diagonal element $H_{l,l}$ represents the $l$-th direct path and the
 off-diagonal elements $H_{l,m}$ for $m\neq l$ represents the FEXT coupling coefficients
 between lines $l$ and $m$.

\begin{figure*}[bp!]
\vspace*{-3mm}
\begin{center}
 \includegraphics[width=0.98\textwidth,angle=0]{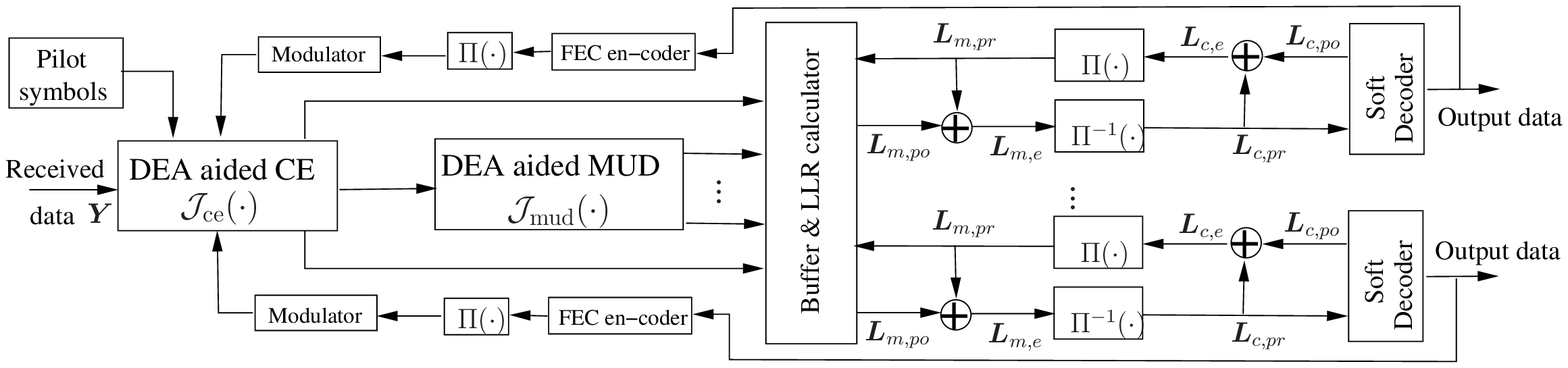}
\end{center}
\vspace{-2mm}
\caption{\small DEA assisted turbo CE and MUD for upstream telecommunications over DSL. The
 subscripts $m$ and $c$ of $\boldsymbol{L}$ are associated with the MUD and channel decoder,
 respectively, while the subscripts $pr$, $po$ and $e$ are used for representing the
 {\it{a priori, a posteriori}} and extrinsic information, respectively.}
\label{FIG2}
\vspace*{-1mm}
\end{figure*}

 In traditional DSL systems, including VDSL2 \cite{eriksson2006vdsl2}, the receiver only
 utilizes the $l$-th entry of $\boldsymbol{Y}$ for the direct-path CE and for the detection
 of the $l$-th user's data. This single-line based detection, i.e., single-user detection
 (SUD), works well for the low-frequency bands, since the magnitudes of the FEXT
 coefficients are much smaller than that of the direct path, which is also known as the
 `diagonal-dominant property' \cite{zafaruddin2015performance}. At the BT Lab at Ipswich,
 U.K., we have measured the frequency-domain channel responses of 100\,m and 200\,m BT
 cables, consisting 10 twisted copper pairs and each wire having a diameter of 0.5\,mm. The
 results are depicted in Fig.~\ref{FIG1}. The VDSL standard \cite{cendrillon2006near} uses
 the frequency band spanning from 25\,kHz to 12\,MHz. As seen from Fig.~\ref{FIG1}\,(a),
 in this frequency band, the FEXT effects are negligible, therefore a SUD is adequate. For
 higher frequency bands of up to 100\,MHz, the `diagonal-dominant property'  of the
 channel matrix remains valid, but the FEXT effects become non-negligible, as shown in
 Fig.~\ref{FIG1}\,(a). Thus, in the upstream transmission, the linear ZF based MUD (ZF-MUD),
 i.e., the ZF-FEXT canceller, can be invoked for removing the interference imposed by
 adjacent lines, which is formulated as 
\begin{align}\label{eq2}
 \hat{\boldsymbol{X}} =& \boldsymbol{H}^{-1}\boldsymbol{Y} = \boldsymbol{X} + \tilde{\boldsymbol{W}} , 
\end{align}
 where $ \tilde{\boldsymbol{W}}=\boldsymbol{H}^{-1}\boldsymbol{W}$ denotes the noise at
 the  output of the ZF-FEXT canceller.

 It can be observed from (\ref{eq2}) that the efficiency of the ZF-MUD relies on the
 diagonal-dominant property. When the ratio of the direct channel magnitude to the FEXT
 interference channel magnitude is high, i.e., $\boldsymbol{H}$ is well-conditioned, the
 inversion $\boldsymbol{H}^{-1}$ is well defined and, therefore, the ZF-FEXT canceller
 efficiently mitigates the interference. Observe from Fig.~\ref{FIG1}\,(a) that the
 higher the operational frequency, the less pronounced this diagonal-dominant property
 becomes, consequently the more seriously the ZF-MUD suffers from noise enhancement, as 
 seen from Fig.~\ref{FIG1}\,(b). Again, the emerging G.fast system will expand the
 bandwidth up to 212\,MHz \cite{effenberger2016future}, where we observe from Fig.~\ref{FIG1}
 that  the direct channels are overwhelmed by the crosstalk. Hence $\boldsymbol{H}$ is
 extremely poorly conditioned and consequently the  ZF-FEXT canceller suffers from an
 extremely high noise enhancement, such as 60\,dB. This motivates our research of more
 powerful MUDs.

\section{DEA Assisted Turbo Channel Estimation and Multi-User Detection}\label{S3}

\subsection{Turbo channel estimation and multi-user detection}\label{S3.1}

 At the $l$-th user's transmitter, where $1\le l\le L$, the bit sequence is first encoded
 by a forward error correction (FEC) code encoder. After passing through an interleaver $\Pi$,
 the coded bit sequence is mapped by an $M$-ary modulator relying on the modulation constellation
 $\mathbb{M}$ into the symbol sequence, which is then transmitted upstream. At the CO receiver,
 the task is to jointly estimate the channel $\boldsymbol{H}$ and to detect the data
 $\boldsymbol{X}$ based on the noisy received signal $\boldsymbol{Y}$. Thus the objective or
 cost function (CF) of this joint CE and MUD is the log likelihood function of $\boldsymbol{Y}$
 conditioned both on $\boldsymbol{H}$ and $\boldsymbol{X}$. Since the noise $\boldsymbol{W}$ is
 white Gaussian, this CF is given by
\begin{align}\label{eq3}
 \mathcal{J}\left(\boldsymbol{H},\boldsymbol{X}\right) =& \left\|\boldsymbol{Y} -
 \boldsymbol{H}\boldsymbol{X}\right\|^{2} .
\end{align}
 The joint ML CE and MUD solution in theory can be found by solving the optimization
 problem:
\begin{align}\label{EQ6:ML} % eq 4
 \left(\hat{\boldsymbol{H}}, \hat{\boldsymbol{X}}\right)^{\star} =& \arg
  \min_{\left( \boldsymbol{H}\in \mathbb{C}^{L \times L},\boldsymbol{X}\in \mathbb{M}^L\right)}
  \mathcal{J}\left(\boldsymbol{H},\boldsymbol{X}\right) ,
\end{align}
 which is unattainable owing to the need of jointly searching the high-dimensional
 continuous channel space and the high-dimensional discrete data space. Note that in the 
 upstream of a DSL system typically dozens of users are served and each user employs an
 $M$-ary modulator. 

 A straightforward suboptimal approach, which is widely adopted in practice, is to first
 estimate the CSI given the training pilots and then to detect the data using the estimated
 channel. To acquire  an adequately accurate CE, however, the number of training pilots must
 be sufficiently large. This approach is therefore inherently suboptimal and significantly
 reduces the achievable throughput. A much better approach is to decompose the
 computationally prohibitive joint ML CE and MUD optimization problem into an iterative CE
 and MUD optimization by using a powerful turbo technique for attaining an iterative gain,
 which is capable of reducing the pilot overhead, while still attaining the optimal
 performance \cite{Turbo6,Turbo7,Turbo8}. Specifically, the joint optimization problem
 (\ref{EQ6:ML}) is solved by the iterative procedure formulated as
\begin{align}\label{eq5}
 \left(\hat{\boldsymbol{H}}, \hat{\boldsymbol{X}}\right)\!  =& \arg
  \min_{\boldsymbol{X}\in\mathbb{M}^L} \mathcal{J}_{\text{mud}}\! \left(\! \boldsymbol{X}\Big|\arg
  \min_{ \boldsymbol{H}\in \mathbb{C}^{L \times L}} \mathcal{J}_{\text{ce}}\left(\boldsymbol{H}\big|
  \breve{\boldsymbol{X}}\right) \! \right) \!\! ,\!
\end{align}
 where the `inner'-optimization performs the CE conditioned on the available data
 $\breve{\boldsymbol{X}}$, which has the CF 
\begin{align}\label{EQ8:Jce} % eq6 
 \mathcal{J}_{\text{ce}}\left(\boldsymbol{H}\big|\breve{\boldsymbol{X}}\right) =&  \left\|\boldsymbol{Y}
  - \boldsymbol{H}\breve{\boldsymbol{X}}\right\|^{2} ,
\end{align}
 while the `outer'-optimization carries out MUD conditioned on the available CSI estimate
 $\breve{\boldsymbol{H}}$, which has the CF 
\begin{align}\label{EQ9:Jmud} % eq7
 \mathcal{J}_{\text{mud}}\left(\boldsymbol{X}\big| \breve{\boldsymbol{H}}\right) =&  \left\|\boldsymbol{Y}
  - \breve{\boldsymbol{H}}\boldsymbol{X}\right\|^{2} .
\end{align}
 The schematic of the proposed turbo CE and MUD procedure  solving the iterative
 optimization (\ref{eq5}) is illustrated in Fig.~\ref{FIG2}.

 Let us denote the iteration index by the superscript $^{(i)}$. In the first iteration, the 
 available data $\breve{\boldsymbol{X}}^{(0)}$ represents the pilot symbols allocated by the
 system. The $i$-th iteration starts by performing the CE: 
\begin{align}\label{EQ11:DE-CE} % eq8
 \breve{\boldsymbol{H}}^{(i)} =& \arg \min_{ \boldsymbol{H}\in \mathbb{C}^{L \times L}}
  \mathcal{J}_{\text{ce}}\left(\boldsymbol{H}|\breve{\boldsymbol{X}}^{(i-1)}\right) ,
\end{align}
 followed by the MUD: 
\begin{align}\label{EQ12:DE-MUD} % eq9
 \tilde{\boldsymbol{X}} =&  \arg \min_{\boldsymbol{X}\in\mathbb{M}^L}
  \mathcal{J}_{\text{mud}}\left(\boldsymbol{X}\Big|\breve{\boldsymbol{H}}^{(i)}\right) .
\end{align}
 Then soft decoding takes place by iteratively exchanging  soft extrinsic information
 between the  MUD and the soft channel decoder of Fig.~\ref{FIG2}.

 Specifically, each detected data symbol $\tilde{X}$ of the $l$-th user, where the
 user index $l$ is omitted for simplicity, is converted into  log likelihood ratios
 (LLRs) by a soft demapper \cite{SoftLLR1}, denoted by $\boldsymbol{L}_{m,po}$, which
 represents the \emph{a posteriori} soft encoded bit information calculated by the soft
 MUD. After subtracting the \emph{a priori} information $\boldsymbol{L}_{m,pr}$ of the
 encoded bits, the extrinsic information delivered by the soft MUD is formulated as
\begin{align}\label{eq10}
 \boldsymbol{L}_{m,e} =& \boldsymbol{L}_{m,po} - \boldsymbol{L}_{m,pr} .
\end{align}
 This is passed through the de-interleaver $\Pi^{-1}$, which becomes the \emph{a priori}
 soft information $\boldsymbol{L}_{c,pr}$ entered into the soft decoder. The decoder
 then decodes  $\boldsymbol{L}_{c,pr}$ to provide the \emph{a posteriori} soft information
 $\boldsymbol{L}_{c,po}$ for the decoded bits. The resultant extrinsic information provided
 by the soft decoder
\begin{align}\label{eq11}
 \boldsymbol{L}_{c,e} =& \boldsymbol{L}_{c,po} - \boldsymbol{L}_{c,pr} ,
\end{align}
 is then passed through the interleaver $\Pi$, and becomes the new \emph{a priori}
 information of the encoded bits. The iterative soft de-mapping and decoding continues until
 the process converges, typically after a few iterations. After the convergence of the soft
 MUD/decoding process, the decoder outputs the hard bits, and this iterative detection/decoding
 is denoted by $\mathcal{C}\left(\tilde{\boldsymbol{X}}\right)$.  The decoded hard bits
 are re-encoded and re-modulated into the data symbols
\begin{align}\label{eq12}
 \breve{\boldsymbol{X}}^{(i)} =& \mathcal{M}\left(\mathcal{C}\left(\tilde{\boldsymbol{X}}\right) \right), 
\end{align}
 which becomes the data available for the next iteration between the CE and MUD. Since 
 the soft channel decoder is capable of producing a reliable bit steam after the convergence
 of the soft MUD/decoder, $\breve{\boldsymbol{X}}^{(i)}$ represents `virtual pilot symbols',
 and this iteration gain of the soft channel decoder will be fully exploited by the CE to
 deliver a more accurate channel estimate $\breve{\boldsymbol{H}}^{(i+1)}$, which in turn
 generates an even more reliable $\breve{\boldsymbol{X}}^{(i+1)}$. The iteration gain of this
 turbo CE and MUD process allows us to gradually approach the  optimal solution of (\ref{EQ6:ML}).

\begin{figure*}[tbp!]
\vspace{-1mm}
\begin{center}
 \includegraphics[width=0.7\linewidth,angle=0]{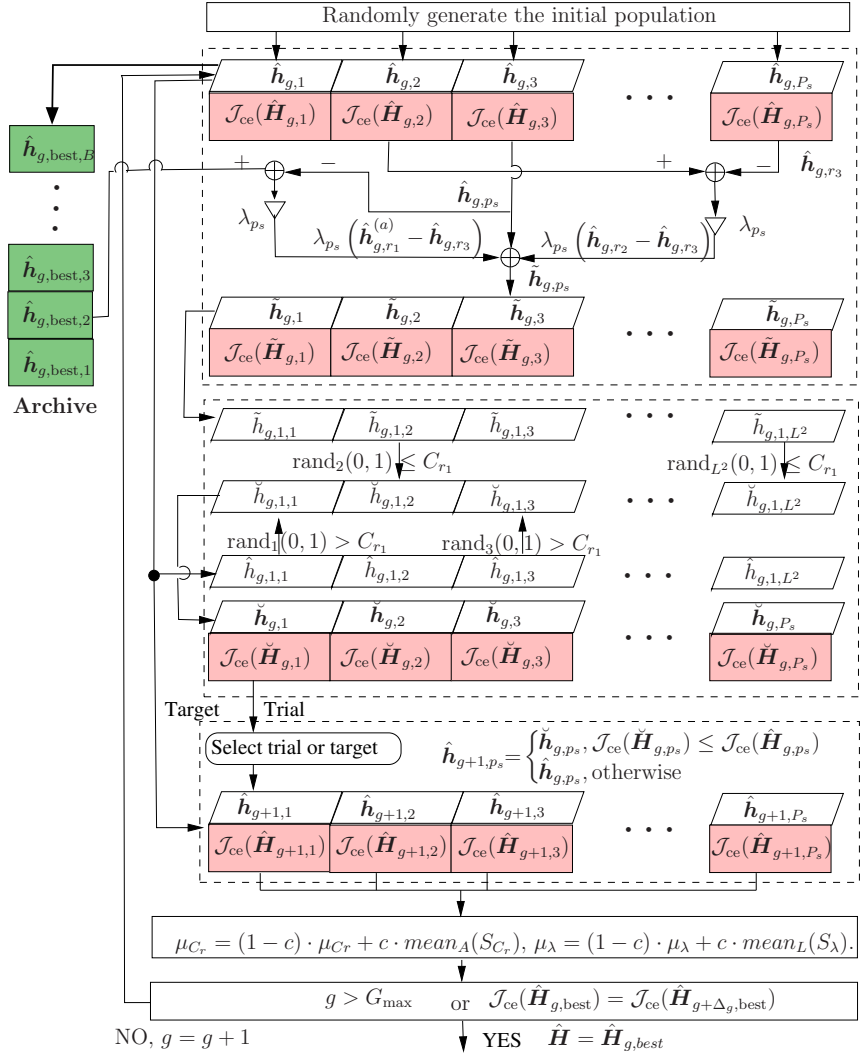}
\end{center}
\vspace{-4mm}
\caption{\small Flowchart of the continuous DEA assisted channel estimation.}
\label{FIG3}
\vspace{-4mm}
\end{figure*}

\subsection{Continuous DEA assisted channel estimation}\label{S3.2}

 The optimization (\ref{EQ11:DE-CE}) of searching the high-dimensional channel space to find
 the optimal $\breve{\boldsymbol{H}}^{(i)}$ can be efficiently carried out by the continuous DEA.
 We now elaborate on this continuous DEA aided CE, whose flowchart is shown in Fig.~\ref{FIG3}.
 For notational convenience, we stack the columns of $\boldsymbol{H}\in \mathbb{C}^{L\times L}$
 and convert it into a vector $\boldsymbol{h}\in \mathbb{C}^{L^2}$.

\begin{enumerate}
\item{\textbf{Initialization}}. At the first generation $g=1$, the initial population of 
 $P_s$ members ${\hat{\boldsymbol{h}}}_{g,p_s}\in \mathbb{C}^{L^2}$ for $1\le p_s\le P_s$
 is randomly and uniformly generated. The mean value of the crossover probability $C_r$ is
 initialized to $\mu_{C_r}=0.5$, while the location parameter of the scaling factor $\lambda$
 is initialized to $\mu_{\lambda}=0.5$. The archive that preserves the $B$ best population
 members is initialized to be empty, where $B=p P_s$ and $0 < p < 1$ is the greedy factor. The
 archive is introduced for preserving the best `genes' of the population.

\item{\textbf{Mutation}}. Each individual ${\hat{\boldsymbol{h}}}_{g,p_s}$, $1\le p_s\le P_s$,
 has the CF value $\mathcal{J}_{ce}\big({\hat{\boldsymbol{H}}}_{g,p_s}\big)$ calculated using
 (\ref{EQ8:Jce}), where ${\hat{\boldsymbol{H}}}_{g,p_s}$ is the channel matrix corresponding
 to ${\hat{\boldsymbol{h}}}_{g,p_s}$. Each population member ${\hat{\boldsymbol{h}}}_{g,p_s}$
 is mutated by adding two scaled difference-vectors, namely ${\hat{\boldsymbol{h}}}_{g,best,r_1}-
 {\hat{\boldsymbol{h}}}_{g,p_s}$ and ${\hat{\boldsymbol{h}}}_{g,r_2}-
 {\hat{\boldsymbol{h}}}_{g,r_3}$, to it  
\begin{align}\label{EQ18:continuous DEmutate} % eq13
 {\tilde{\boldsymbol{h}}}_{g,p_s} =& {\hat{\boldsymbol{h}}}_{g,p_s} +
  \lambda_{p_s}\big({\hat{\boldsymbol{h}}}_{g,best,r_1} - {\hat{\boldsymbol{h}}}_{g,p_s}\big) \nonumber\\ &
  + \lambda_{p_s}\left({\hat{\boldsymbol{h}}}_{g,r_2} - {\hat{\boldsymbol{h}}}_{g,r_3}\right) ,
\end{align}
 where ${\hat{\boldsymbol{h}}}_{g,best,r_1}$ is randomly selected from the archive, i.e. 
 $r_1$ is randomly selected from $\{1,2,\cdots ,B\}$, $r_2$ and $r_3$ are two values randomly
 selected from $\{1,2,\cdots ,(p_s -1),(p_s + 1),\cdots ,P_s\}$, while $\lambda_{p_s}\in
 (0, ~ 1]$ is a randomly generated scaling factor according to the following procedure. Draw
 a random number $\gamma$ according to the Cauchy distribution \cite{Johnson1994continuous}
 with the location parameter $\mu_{\lambda}$ and the scale parameter $\sigma_{\lambda}$: If
 $\gamma\le 0$, re-draw $\gamma$; if $\gamma\in (0, ~ 1]$, then set $\lambda_{p_s}=\gamma$; if $\gamma > 1$,
 then use $\lambda_{p_s}=1$. The `mutated' individual ${\tilde{\boldsymbol{h}}}_{g,p_s}$, $1\le p_s
 \le P_s$, has the CF value $\mathcal{J}_{ce}\big({\tilde{\boldsymbol{H}}}_{g,p_s}\big)$,
 where ${\tilde{\boldsymbol{H}}}_{g,p_s}$ is the channel matrix corresponding to
 ${\tilde{\boldsymbol{h}}}_{g,p_s}$.

\item{\bf{Crossover}}. The continuous DEA generates a `trial' individual $\breve{\boldsymbol{h}}_{g,p_s}$
 by exchanging some elements of the `target' individual ${\hat{\boldsymbol{h}}}_{g,p_s}$
 with the corresponding elements of the `donor' individual ${\tilde{\boldsymbol{h}}}_{g,p_s}$.
 Explicitly, the crossover operation on the $\alpha$-th element is given by
\begin{align}\label{EQ19:crossover} % eq14
 \breve{h}_{g,p_s,\alpha} =& \left\{\begin{array}{ll} \tilde{h}_{g,p_s,\alpha},
  & \text{rand}_{\alpha}(0,1) \le C_{r_{p_s}} , \\
 \hat{h}_{g,p_s,\alpha}, & \text{otherwise}, \end{array}\right.
\end{align}
 where $\text{rand}_{\alpha}(0,1)$ denotes the random number drawn from the uniform
 distribution in $[0, ~ 1]$ for the $\alpha$-th element, while $C_{r_{p_s}}\in [0,~ 1]$ is
 the crossover probability, which is randomly generated according to the following procedure.
 Draw a random number $\gamma$ according to the normal distribution with the mean $\mu_{C_r}$
 and the standard deviation $\sigma_{C_r}$: If $\gamma < 0$, re-draw $\gamma$; if $\gamma\in
 [0, ~ 1]$, then set $C_{r_{p_s}}=\gamma$; if $\gamma > 1$, then use $C_{r_{p_s}}=1$. In
 Fig.~\ref{FIG3}, this crossover operation for $p_s=1$ is illustrated. The trial individual
 $\breve{\boldsymbol{h}}_{g,p_s}$, $1\le p_s\le P_s$, has the CF value
 $\mathcal{J}_{\text{ce}}\big(\breve{\boldsymbol{H}}_{g,p_s}\big)$, where again the matrix
 $\breve{\boldsymbol{H}}_{g,p_s}$ corresponds to $\breve{\boldsymbol{h}}_{g,p_s}$.

\item{\bf{Selection}}. The selection operation decides whether the target vector
 ${\hat{\boldsymbol{h}}}_{g,p_s}$ or the trial vector $\breve{\boldsymbol{h}}_{g,p_s}$
 will survive to the next generation according to their CF values
\begin{align}\label{EQ20:select} % eq15
 {\hat{\boldsymbol{h}}}_{g+1,p_s}\! =& \! \left\{ \! \begin{array}{cl} \breve{\boldsymbol{h}}_{g,p_s},
 \!\! &\!\! \mathcal{J}_{\text{ce}}\big(\breve{\boldsymbol{H}}_{g,p_s}\big) \le
  \mathcal{J}_{\text{ce}}\big({\hat{\boldsymbol{H}}}_{g,p_s}\big) , \\
 {\hat{\boldsymbol{h}}}_{g,p_s},\!\! &\!\! \text{otherwise} .\end{array}\right.\!\!\!
\end{align}
 The archive is replaced by the $100 p P_s\%$ best individuals of the new population
 $\left\{{\hat{\boldsymbol{h}}}_{g+1,p_s}\right\}_{p_s=1}^{P_s}$.

\item{\bf{Adaptation}}. To keep up with the `evolution', the mean of the crossover
 probability $\mu_{C_r}$ and the location parameter of the scaling factor $\mu_{\lambda}$
 are adaptively updated according to
\begin{align}
 \mu_{C_r} =& (1-c) \cdot \mu_{C_r} + c \cdot mean_{A}(\mathcal{S}_{C_r}), \label{eq16} \\
 \mu_{\lambda} =& (1-c) \cdot \mu_{\lambda} + c \cdot
 mean_{L}(\mathcal{S}_{\lambda}) , \label{eq17}
\end{align}
 where $c \in (0, ~ 1]$ is the adaptive update factor controlling the rate of the
 parameter adaptation, $mean_{A}(\cdot )$ and $mean_{L}(\cdot )$ denote the
 arithmetic-mean and Lehmer-mean \cite{Havil2003gamma} operators, respectively, while
 $\mathcal{S}_{C_r}$ and $\mathcal{S}_{\lambda}$ denote the sets of the successful
 crossover probabilities $C_{r_{p_s}}$ and scaling factors $\lambda_{p_s}$ of generation
 $g$, respectively.
	
\item{\bf{Termination}}. The ideal stopping criterion would be the convergence of
 the population. In practice, we opt for halting the optimization procedure, when any of 
 the following two stopping criteria is met:
 \begin{itemize}
 \item The pre-set maximum number of generations $G_{\text{max}}$ has been exhausted.
 \item $\Delta_g$ generations have been explored without any reduction in the CF value
  associated with the best individual in the population.
 \end{itemize}
 Otherwise, set $g=g+1$, and go to 2)~\textbf{Mutation}.
\end{enumerate}

\begin{figure*}[bp!]
\vspace{-2mm}
\begin{center}
 \includegraphics[width=0.7\linewidth,angle=0]{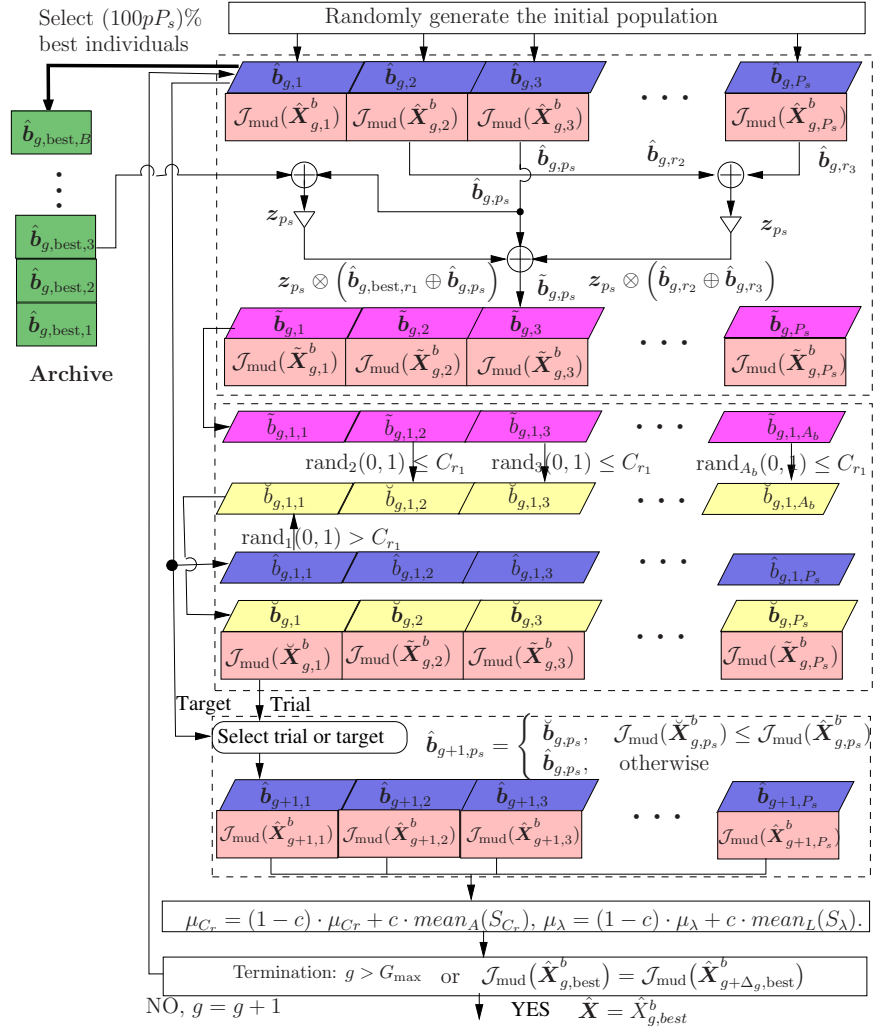}
\end{center}
\vspace{-4mm}
\caption{\small Flowchart of the discrete DEA assisted multi-user detection.}
\label{FIG4}
\vspace{-1mm}
\end{figure*}

 The scale parameter $\sigma_{\lambda}$ of the scaling factor and the standard deviation
 $\sigma_{C_r}$ of the crossover probability should be set to a small value, e.g.,
 $\sigma_{\lambda}=0.1$ and $\sigma_{C_r}=0.1$. The remaining algorithmic parameters 
 to be set are the population size $P_s$, the greedy factor $p$, the adaptive update factor
 $c$, the maximum number of generations $G_{\text{max}}$ and/or the value of $\Delta_g$
 for terminating the continuous DEA aided CE.

\subsection{Discrete DEA aided multi-user detection}\label{S3.3}

 Clearly, it is impractical to find the optimal ML solution ${\tilde{\boldsymbol{X}}}$ for
 the optimization (\ref{EQ12:DE-MUD}) by exhaustive search, when the number of upstream users
 served $L$ is large and/or the modulation order $M$ is very high. However, the optimization
 (\ref{EQ12:DE-MUD}) of searching the high-dimensional data space to find the optimal
 ${\tilde{\boldsymbol{X}}}$ can be carried out by a discrete DEA, at a fraction of the
 computational complexity imposed by the exhaustive-search based optimal ML-MUD. This discrete
 DEA assisted MUD is depicted in Fig.~\ref{FIG4}. We will denote the bit vector mapped to the
 symbol vector
 $\boldsymbol{X}$ by $\boldsymbol{b}=\big[b_1 ~ b_2 \cdots b_{A_b}\big]^{\rm T}$, where each
 bit $b_i$ takes the value of 1 or 0, and $A_b=L\cdot \log_2 M$.
 
\begin{enumerate}
\item{\bf{Initialization.}}
 At the first generation of $g=1$, the discrete DEA randomly initializes its population of $P_s$
 individuals
\begin{align}\label{eq18}
 \hspace*{-2mm}\hat{\boldsymbol{b}}_{g,p_s}\! =& \Big[\hat{b}_{g,p_s,1} ~ \hat{b}_{g,p_s,2}
  \cdots \hat{b}_{g,p_s,A_b}\Big]^{\rm T} , 1\le p_s\le P_s , \!
\end{align}
 i.e., every bit $\hat{b}_{g,p_s,i}$ is randomly assigned  1 or 0. The mean of the
 crossover probability $C_r$ and the location parameter of the scaling factor $\lambda$ are 
 initialized to $\mu_{C_r}=0.5$ and $\mu_{\lambda}=0.5$, respectively. The archive that
 preserves  the $B$ best individuals of the previous generation is set to empty.

\item{\bf{Mutation.}} Each individual $\hat{\boldsymbol{b}}_{g,p_s}$, $1\le p_s\le P_s$,
 corresponds to a modulated symbol vector ${ \hat{\boldsymbol{X}}}_{g,p_s}^b$ associated
 with the CF value $\mathcal{J}_{mud}\big({ \hat{\boldsymbol{X}}}_{g,p_s}^b\big)$ that is
 calculated using (\ref{EQ9:Jmud}). The discrete DEA mutates each base population vector
 $\hat{\boldsymbol{b}}_{g,p_s}$ by adding two appropriately scaled and randomly selected
 difference-vectors to it. Note that in the binary arithmetic, the scaling or multiplying
 operation is represented by the bit-wise exclusive-AND operator $\otimes$, while the
 addition or difference operation is represented by the bit-wise exclusive-OR operator $\oplus$.
 Explicitly, the `mutated' individual is given by
\begin{align} \label{EQ24:DE_mutation} % eq19
 \tilde{\boldsymbol{b}}_{g,p_s} =& \hat{\boldsymbol{b}}_{g,p_s} \oplus \left(
  \boldsymbol{z}^b_{p_s}\otimes \left(\hat{\boldsymbol{b}}_{g,best,r_1} \oplus
  \hat{\boldsymbol{b}}_{g,p_s}\right)\right) \nonumber \\
	& \oplus \left( \boldsymbol{z}^b_{p_s} \otimes
  \left(\hat{\boldsymbol{b}}_{g,r_2} \oplus \hat{\boldsymbol{b}}_{g,r_3}\right)\right) ,
\end{align}
 where $\hat{\boldsymbol{b}}_{g,best,r_1}$ is randomly selected from the archive,
 $\hat{\boldsymbol{b}}_{g,r_2}$ and $\hat{\boldsymbol{b}}_{g,r_3}$ are randomly
 selected from the rest of the current population, while the bit-scaling factor
 $\boldsymbol{z}^b_{p_s}=\big[z^b_{p_s,1} ~ z^b_{p_s,2}\cdots z^b_{p_s,A_b}\big]^{\rm T}$
 is the $A_b$-length binary-valued vector generated randomly according to the following
 procedure. First, the real-valued vector $\boldsymbol{z}_{p_s}=\big[z_{p_s,1} ~ z_{p_s,2}
 \cdots z_{p_s,A_b}\big]^{\rm T}\in\mathbb{R}^{A_b}$ is generated, whose elements all
 obey the Gaussian distribution of zero mean and unity variance. Then the scaling factor
 $\lambda_{p_s}\in (0, ~ 1]$ is generated according to the Cauchy distribution with the 
 location parameter $\mu_{\lambda}$ and the scaling parameter of $\sigma_{\lambda}$,
 similar to the generation of the scaling factor in the mutation step of the continuous
 DEA. By comparing  the elements of $\boldsymbol{z}_{p_s}$ with $\lambda_{p_s}$, the
 elements of  $\boldsymbol{z}^b_{p_s}$ are determined according to
\begin{align}\label{eq20}
 z^b_{p_s,i} =& \left\{ \begin{array}{cl}
  1 , & \text{if } z_{p_s,i} < \lambda_{p_s} , \\ 0 , & \text{otherwise} , \end{array} \right.
\end{align}
 for $1\le i\le A_b$. Each  `mutated' individual $\tilde{\boldsymbol{b}}_{g,p_s}$, $1\le p_s
\le P_s$, has the associated CF value $\mathcal{J}_{mud}\big({\tilde{\boldsymbol{X}}}_{g,p_s}^b\big)$,
 where ${\tilde{\boldsymbol{X}}}_{g,p_s}^b$ is the symbol vector corresponding to the bit
 vector $\tilde{\boldsymbol{b}}_{g,p_s}$.

\item{\bf{Crossover.}} The discrete DEA generates a trial vector $\breve{\boldsymbol{b}}_{g,p_s}$
 by replacing certain elements of the target vector $\hat{\boldsymbol{b}}_{g,p_s}$ with the
 corresponding elements of the donor vector $\tilde{\boldsymbol{b}}_{g,p_s}$. There exist
 diverse variants of this crossover mechanism \cite{price2005differential,qin2009differential},
 and we adopt the uniform crossover algorithm. Specifically, the $\alpha$-th element of
 $\breve{\boldsymbol{b}}_{g,p_s}$ is determined according to
\begin{align}\label{eq21}
 \breve{{b}}_{g,p_s,\alpha} =& \left\{\begin{array}{ll}\tilde{{b}}_{g,p_s,\alpha}, &
  \text{rand}_{\alpha}(0,1) \le C_{r_{p_s}} , \\ 
  \hat{{b}}_{g,p_s,\alpha}, & \text{otherwise} , \end{array}\right.
\end{align}
 where $C_{r_{p_s}}\in [0, ~1]$ is the crossover probability, randomly generated according to
 the normal distribution of mean $\mu_{C_r}$ and standard deviation $\sigma_{C_r}$, similar to
 the generation of the crossover probability in the crossover step of the continuous DEA. This
 crossover operation for $p_s=1$ is depicted in Fig.~\ref{FIG4}. The trial individual
 $\breve{\boldsymbol{b}}_{g,p_s}$, $1\le p_s\le P_s$, has the CF value
 $\mathcal{J}_{mud}\big(\breve{\boldsymbol{X}}_{g,p_s}^b\big)$, where $\breve{\boldsymbol{X}}_{g,p_s}^b$
 denotes the symbol vector corresponding to $\breve{\boldsymbol{b}}_{g,p_s}$.

\item{\bf{Selection.}} 
 Whether the target vector $\hat{\boldsymbol{b}}_{g,p_s}$ or the trial vector
 $\breve{\boldsymbol{b}}_{g,p_s}$ survives into the next generation is decided according
 to their associated CF values. Specifically, for $1\le p_s\le P_s$,
\begin{align}\label{eq22}
 \hat{\boldsymbol{b}}_{g+1,p_s} =& \left\{ \begin{array}{ll}\breve{\boldsymbol{b}}_{g,p_s}, &
  {\cal J}_{\text{mud}}\big(\breve{\boldsymbol{X}}^b_{g,p_s}\big) \le {\cal J}_{\text{mud}}\big({ \hat{\boldsymbol{X}}}^b_{g,p_s}),  \\
  \hat{\boldsymbol{b}}_{g,p_s}, & \text{otherwise} .\end{array}\right.
\end{align}
 The archive is replaced by the $B=100 p P_s\%$ best individuals of the new population
 $\left\{\hat{\boldsymbol{b}}_{g+1,p_s}\right\}_{p_s=1}^{P_s}$.

\item{\bf{Adaptation.}} Similar to the continuous DEA, the mean of the crossover probability
 $\mu_{C_r}$ and the location parameter of the scaling factor $\mu_{\lambda}$ are updated
 according to \cite{qin2009differential}
\begin{align} % eq20 - eq21
 \mu_{C_r} =& (1-c) \cdot \mu_{C_{r}} + c \cdot {mean}_A\big(\mathcal{S}_{C_r}\big) , \label{eq23} \\
 \mu_{\lambda} =& (1-c) \cdot \mu_{\lambda} + c \cdot {mean}_L\big(\mathcal{S}_{\lambda}\big) , \label{eq24}
\end{align}
 where again $c\in (0,~ 1]$ is the adaptive update factor, $mean_A(\cdot )$ and 
 $mean_{L}(\cdot )$ denotes the arithmetic mean and Lehmer mean, respectively, while
 $\mathcal{S}_{C_r}$ and $\mathcal{S}_{\lambda}$ denote the sets of the successful
 crossover probabilities $C_{r_{p_s}}$ and scaling factors $\lambda_{p_s}$ of generation $g$,
 respectively.

\item{\bf{Termination.}}
 The optimization procedure is halted when any of the following two stopping criteria are met:
 \begin{itemize}
 \item The pre-set maximum number of generations $G_{\text{max}}$ has been exhausted.
 \item $\Delta_g$ generations have been explored without any reduction in the CF value
  associated with the best individual in the population.
 \end{itemize}
 Otherwise, set $g=g+1$, and go to 2)~\textbf{Mutation}.
\end{enumerate}

\begin{table*}[tp!]
{{\scriptsize
\vspace*{-1mm}
\caption{\small Default algorithmic and system parameters}
\vspace*{-2mm}
\label{Table:Sim_para} % TAB1
\begin{center}
\begin{tabular}{||c|l|c||c|l|c||c|l|c||}
\hline\hline  
  & Initialization & Randomly & & Initialization & Randomly& &Channel code & Turbo \\
  & Population size $P_s$ & 100 & & Population size $P_s$ & 100 & & Code rate & 1/2 \\
 DEA assisted & Greedy factor $p$ & 0.1 & DE assisted & Greedy factor $p$ & 0.1 & System & Memory length & 16 \\
 CE & Adaptive factor $c$ & 0.1 & MUD & Adaptive factor $c$ & 0.8 & parameters & Polynomial & $(3, [7,5 ])$ \\
  & $G_{\max}$ & 100 & & $G_{\max}$ & 100 & & Users & 4 \\
  & $\Delta g_{\max}$ & 20  & & $\Delta g_{\max}$ & 20& & Modulation & 16-QAM \\ \hline\hline  
\end{tabular}
\end{center}
\vspace*{-6mm}
}}
\end{table*}

 Similar to the continuous DEA, the scale parameter $\sigma_{\lambda}$ of the scaling
 factor and the standard deviation $\sigma_{C_r}$ of the crossover probability can both
 be set to 0.1. The user has to set the population size $P_s$, the greedy factor $p$,
 the adaptive update factor $c$, the maximum number of generations $G_{\text{max}}$
 and/or the value of $\Delta_g$ for terminating the discrete DEA aided MUD.

{
\section{Cramer-Rao lower bound of channel estimation}\label{S4}
}

 The CRLB provides the lowest achievable mean square error (MSE) of any unbiased estimator
 \cite{kay1993fundamentals}. In the simulation section, we will demonstrate that our DEA
 assisted CE is capable of approaching the CRLB. Therefore, below we derive the CRLB of
 the channel estimator. Since the CRLB is related to the available training symbols, we
 will introduce the symbol index $[s]$. Thus, upon recalling (\ref{EQ1_re_sig}), we rewrite
 the $l$-th received signal at the $s$-th symbol $Y_l[s]$ as 
\begin{align}\label{eq25}
 Y_l[s] =& \boldsymbol{H}_{[l:\,]}\boldsymbol{X}[s] + W_l[s] ,
\end{align} 
 where $\boldsymbol{H}_{[l:\,]}$ is the $l$-th row of the channel matrix $\boldsymbol{H}$,
 $\boldsymbol{X}[s]=\left[ X_1[s] ~ X_2[s]\cdots X_L[s]\right]^{\rm T}$ is the transmitted
 signal vector at the $s$-th symbol, and $W_l[s]$ is the $l$-th element of $\boldsymbol{W}[s]$.

 Since $W_l[s]\in\mathbb{C}$ represents the white Gaussian noise with covariance $\sigma_w^2$, the
 conditional probability density function (PDF) $f\left(Y_l[s]|\boldsymbol{H}_{[l:\,]}\right)$
 is given by
\begin{align}\label{EQ30:pdf} % eq26
 f\left(Y_l[s]|\boldsymbol{H}_{[l:\,]}\right) =& \frac{1}{2\pi\sigma_w^2}\exp
  \left(-\frac{\left| Y_l[s] - \boldsymbol{H}_{[l:\,]}\boldsymbol{X}[s]\right|^2}{2\sigma_{w}^2}\right) .
\end{align}
 Thus, the joint conditional PDF over the $S$ consecutive OFDM symbols,
 $f\left(Y_l[1],Y_l[2],\cdots ,Y_l[S]|\boldsymbol{H}_{[l:\,]}\right)$, can be formulated as
\begin{align}\label{EQ31:pdf} % eq27
 &f\left(Y_l[1],Y_l[2],\cdots ,Y_l[S]|\boldsymbol{H}_{[l:\,]}\right) \nonumber\\
&\qquad= \prod\limits_{s=1}^S
  \left( \frac{1}{2\pi\sigma_w^2}\exp \left(-\frac{\left|Y_l[s] - \boldsymbol{H}_{[l:\,]}
  \boldsymbol{X}[s]\right|^2}{2\sigma_w^2}\right)\right) .
\end{align}
 The Fisher information matrix is defined as \cite{kay1993fundamentals}
\begin{align}\label{EQ32:fisher} % eq28
 \boldsymbol{I}\big(\boldsymbol{H}_{[l:\,]}\big) =& -\mathcal{E}\left\{\frac{\partial^2 \log
  f\left(Y_l[1],Y_l[2],\cdots ,Y_l[S]|\boldsymbol{H}_{[l:\,]}\right)}{\partial \boldsymbol{H}_{[l:\,]}
  \partial \boldsymbol{H}_{[l:\,]}^{\rm H}}\right\} \nonumber\\
	=& \frac{1}{2\sigma_w^2}\sum\limits_{s=1}^S
  \mathcal{E}\left\{\boldsymbol{X}[s]\left(\boldsymbol{X}[s]\right)^{\rm H}\right\} \!\! , \!
\end{align}
 where $\mathcal{E}\left\{ ~ \right\}$ denotes the expectation operator.

 The CRLB for the estimate of $\boldsymbol{H}_{[l:\,]}$ is defined as
\begin{align}\label{eq29}
 \text{CRLB}\left(\boldsymbol{H}_{[l:\,]}\right) &= \text{Tr}\left(\boldsymbol{I}^{-1}
  \big(\boldsymbol{H}_{[l:\,]}\big)\right) \nonumber\\
 & \hspace*{-10mm}= 2\sigma_w^2 \text{Tr}\left( \left( \sum\limits_{s=1}^S
  \mathcal{E}\left\{\boldsymbol{X}[s]\left(\boldsymbol{X}[s]\right)^{\rm H}\right\}\right)^{-1}\right) ,
\end{align}
 where $\text{Tr}( \cdot )$ denotes the matrix trace operation. Since the optimal training symbol
 sequence satisfies $\mathcal{E}\left\{\boldsymbol{X}[s]\left(\boldsymbol{X}[s]\right)^{\rm H}\right\}
 =E_{\rm s}\boldsymbol{I}_L$, where $\boldsymbol{I}_L$ denotes the $L\times L$ identity matrix and 
 $E_{\rm s}$ is the average power of the transmitted data symbol. Thus, the CRLB can be expressed as
\begin{align}\label{eq30}
 \text{CRLB}\left(\boldsymbol{H}_{[l:\,]}\right)  =& \frac{2\sigma_{w}^2}{S E_{\rm s}} .
\end{align}
 Furthermore, the normalized CRLB (NCRLB), which represents the lower-bound of the achievable NMSE,
 is given by
\begin{align}\label{eq31}
 \text{NCRLB}\left(\boldsymbol{H}_{[l:\,]}\right)  =& \frac{2\sigma_w^2}{S E_{\rm s}
  \left\|\boldsymbol{H}_{[l:\,]}\right\|^2} .
\end{align}

\section{Simulation Results}\label{S5}

 In this section, we evaluate the performance of the proposed DEA assisted turbo CE and MUD
 for upstream G.fast systems operated in the frequency range of 2\,MHz to 212\,MHz, which
 are split into 4096 tones. The channel was characterized by the measurements of BT's
 twisted copper lines at the BT Ultra-Fast lab. The number of simultaneous upstream users
 has the default value of $L=4$ and each user employs the same 16-QAM scheme combined with a
 half-rate turbo channel code of memory 16 using the octal generator polynomials of $(3,[7, 5])$.
 The total number of CF evaluations for the ML-MUD is $16^4 = 65536$. The default loop length
 of the DSL lines is 100\,m. The iterative procedures of the inner turbo decoding is
 automatically terminated, when they have converged. The number of iterations between the DEA
 assisted CE and the DEA aided MUD is set to 6 in our investigation. Note that our proposed
 scheme is readily applicable to systems supporting a number of users and higher-order modulation.
 However, since we use the optimal ML solution as the ultimate benchmark of our proposed scheme
 and we can only compute the optimal ML solution for the system supporting a low number of
 users associated with relatively low-order modulation, we restrict our simulation study to
 $L=4$ and 16-QAM modulation. The default algorithmic parameters used for the continuous and
 discrete DEAs and the system parameters are summarized in Table~\ref{Table:Sim_para}. Unless
 otherwise specified, these default parameter values are used throughout.

 We first investigate the per subcarrier performance and the convergence performance of the
 individual continuous DEA aided CE and discrete DEA aided MUD components, respectively, as
 well as the impact of the system bandwidth on the achievable performance. Both the least
 square (LS) CE and the NCRLB are used as the benchmarks for evaluating the continuous DEA
 aided CE (DEA-CE), where the NCRLB indicates the best achievable performance of the channel
 estimator. Furthermore, the SUD, the ZF-MUD and the ML-MUD are used as the benchmarks of the
 discrete DEA aided MUD (DEA-MUD), where the ML-MUD provides the best achievable detection
 performance. Then the performance of the proposed turbo DEA-CE and DEA-MUD is investigated
 for quantifying the achievable iterative gain of this turbo CE and MUD scheme. Furthermore,
 the impact of the impulse noise as well as that of the CE error and the loop length on the
 detection performance are also investigated.
 
\begin{figure*}[tp!]
\vspace{-1mm}
\begin{center}
\includegraphics[width=0.45\textwidth,angle=0]{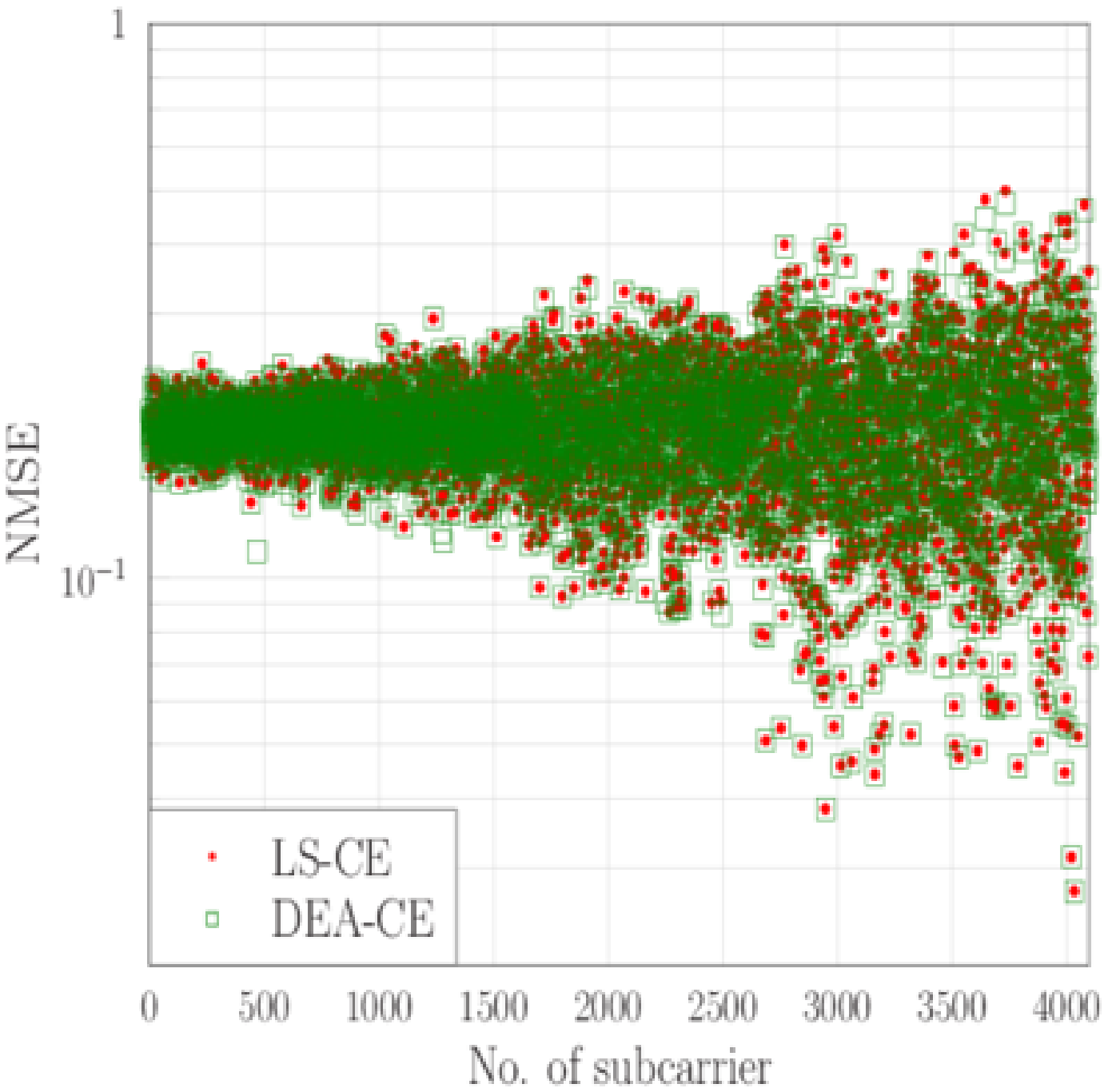}\hspace*{4mm}\includegraphics[width=0.45\textwidth,angle=0]{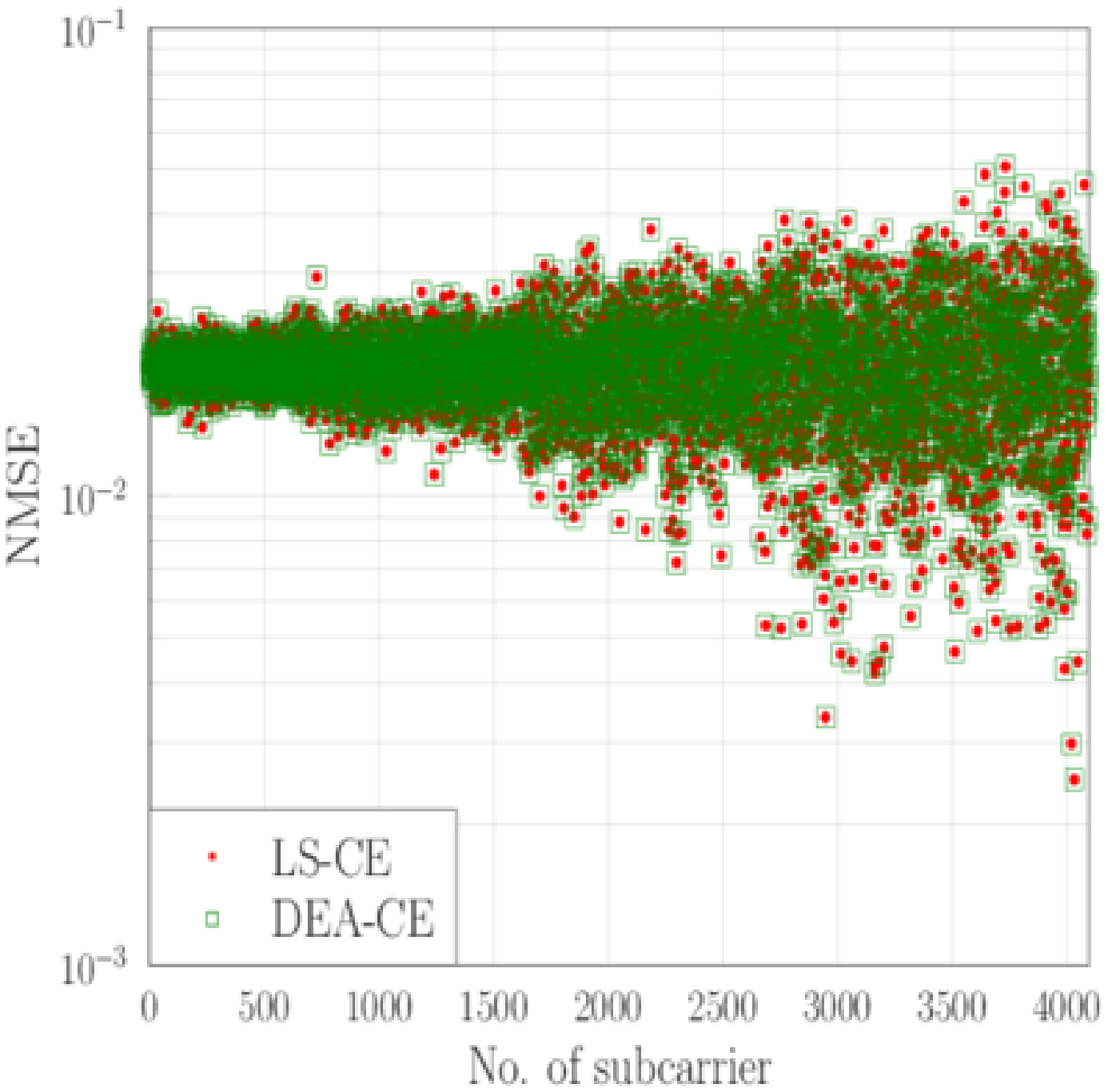}
\end{center}
\vspace{-5mm}
\begin{center}\hspace*{7mm}{\small (a)}\hspace*{80mm}{\small (b)}\end{center}
\vspace{-4mm}
\caption{\small NMSE versus the subcarrier index: (a)~$E_b/N_0=20$\,dB, and (b)~$E_b/N_0=30$\,dB.}
\label{FIG5}
%\vspace{-4mm}
%\end{figure*}
%
%\begin{figure*}[tp!]
%\vspace{-1mm}
\begin{center}
\includegraphics[width=0.45\textwidth,angle=0]{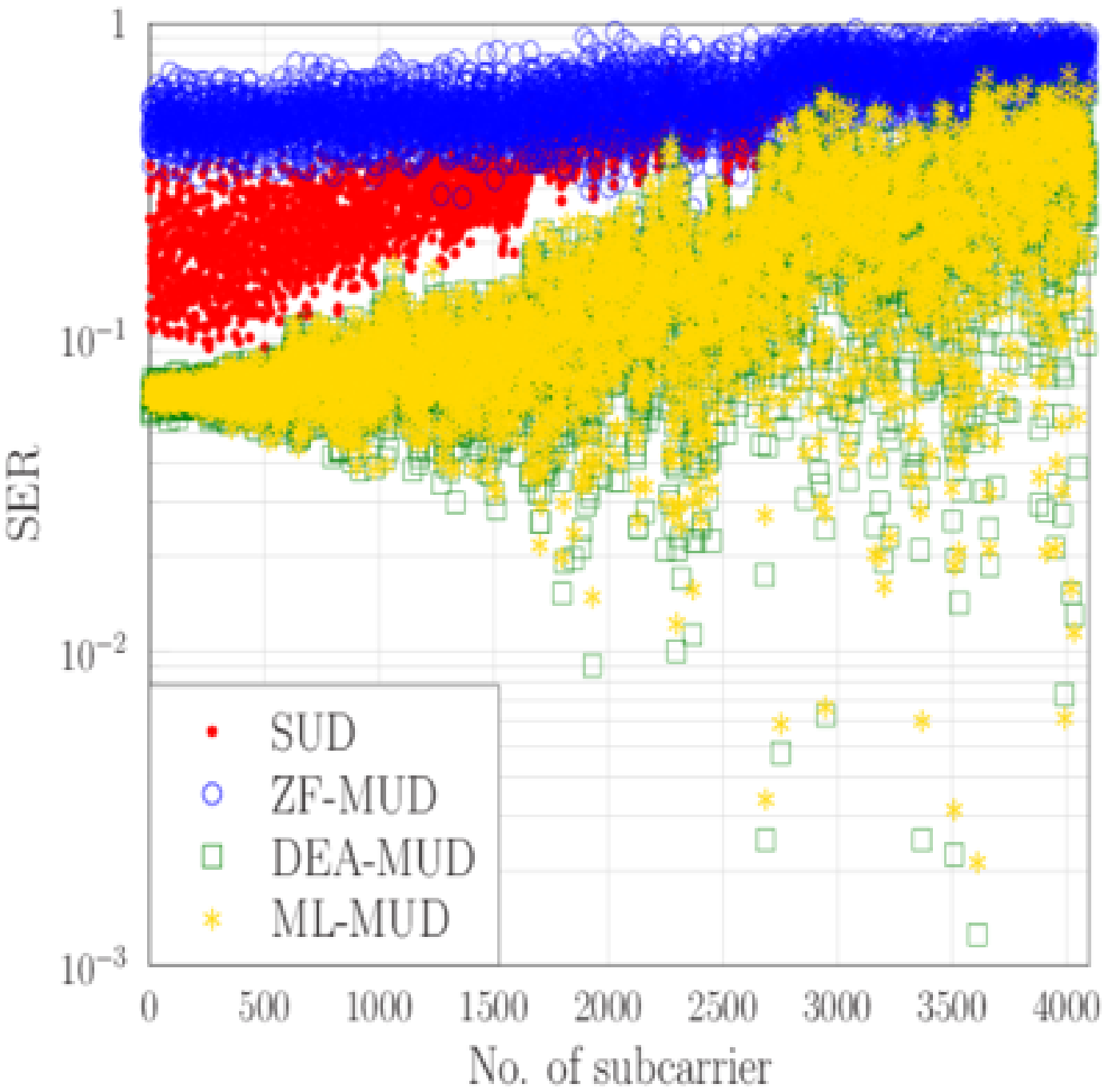}\hspace*{4mm}\includegraphics[width=0.45\textwidth,angle=0]{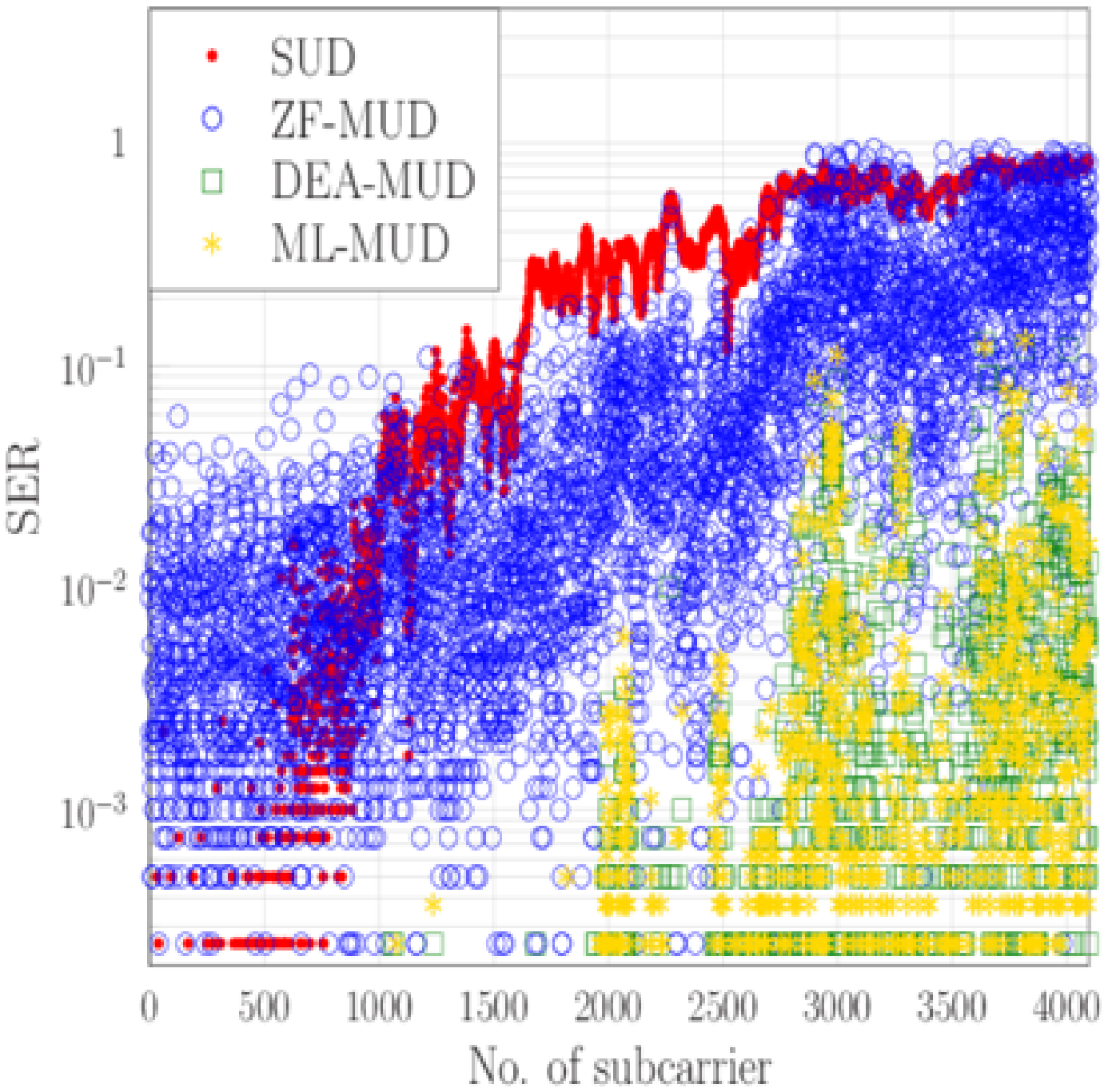}
\end{center}
\vspace{-5mm}
\begin{center}\hspace*{7mm}{\small (a)}\hspace*{80mm}{\small (b)}\end{center}
\vspace{-4mm}
\caption{\small Idealized SER based on perfect CSI versus the subcarrier index: (a)~$E_b/N_0=20$\,dB,
 and (b)~$E_b/N_0=30$\,dB.}
\label{FIG6}
\vspace{-4mm}
\end{figure*}

\begin{figure*}[bp!]
\vspace{-4mm}
\begin{center}
\includegraphics[width=0.45\textwidth,angle=0]{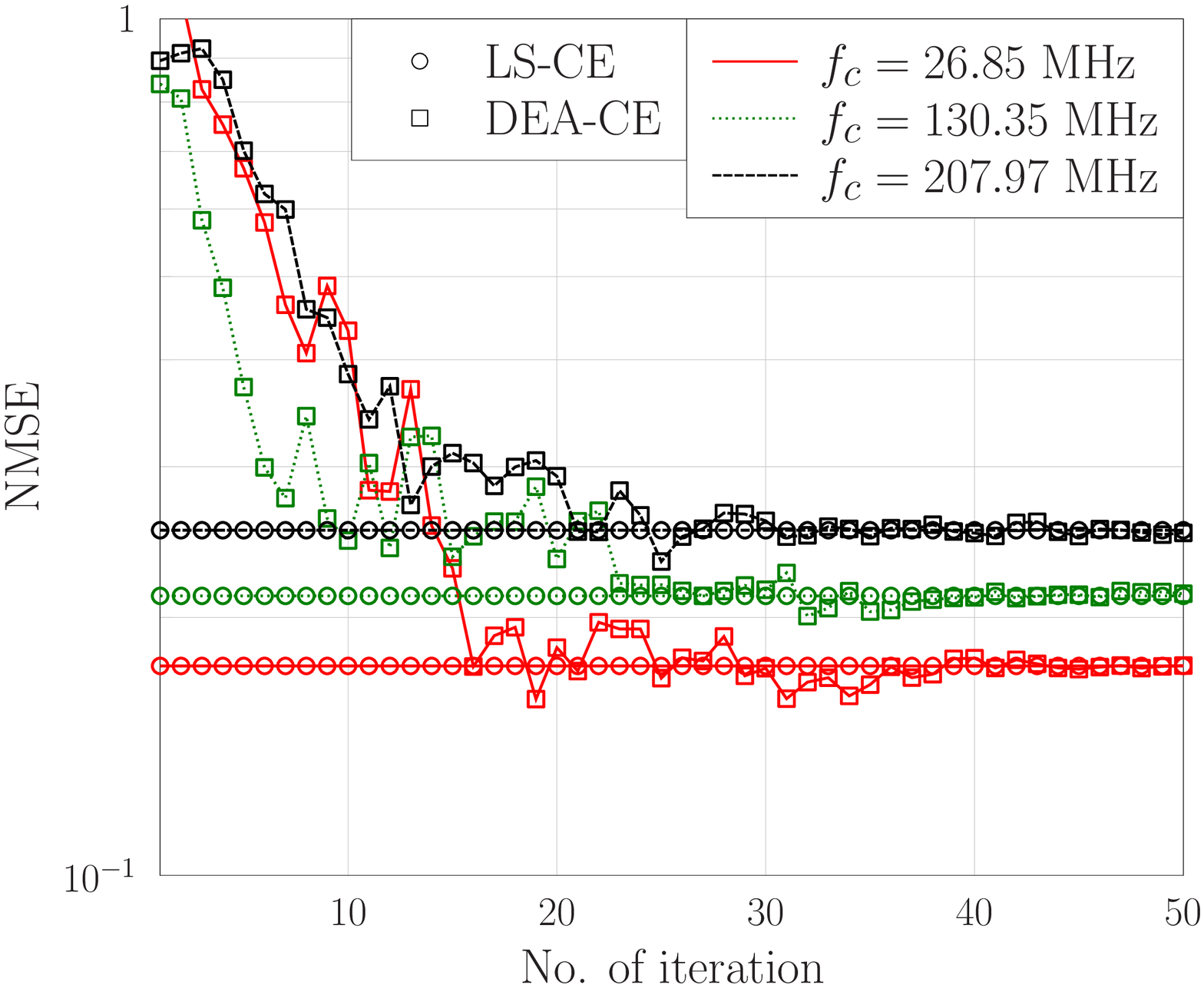}\hspace*{2mm}\includegraphics[width=0.45\textwidth,angle=0]{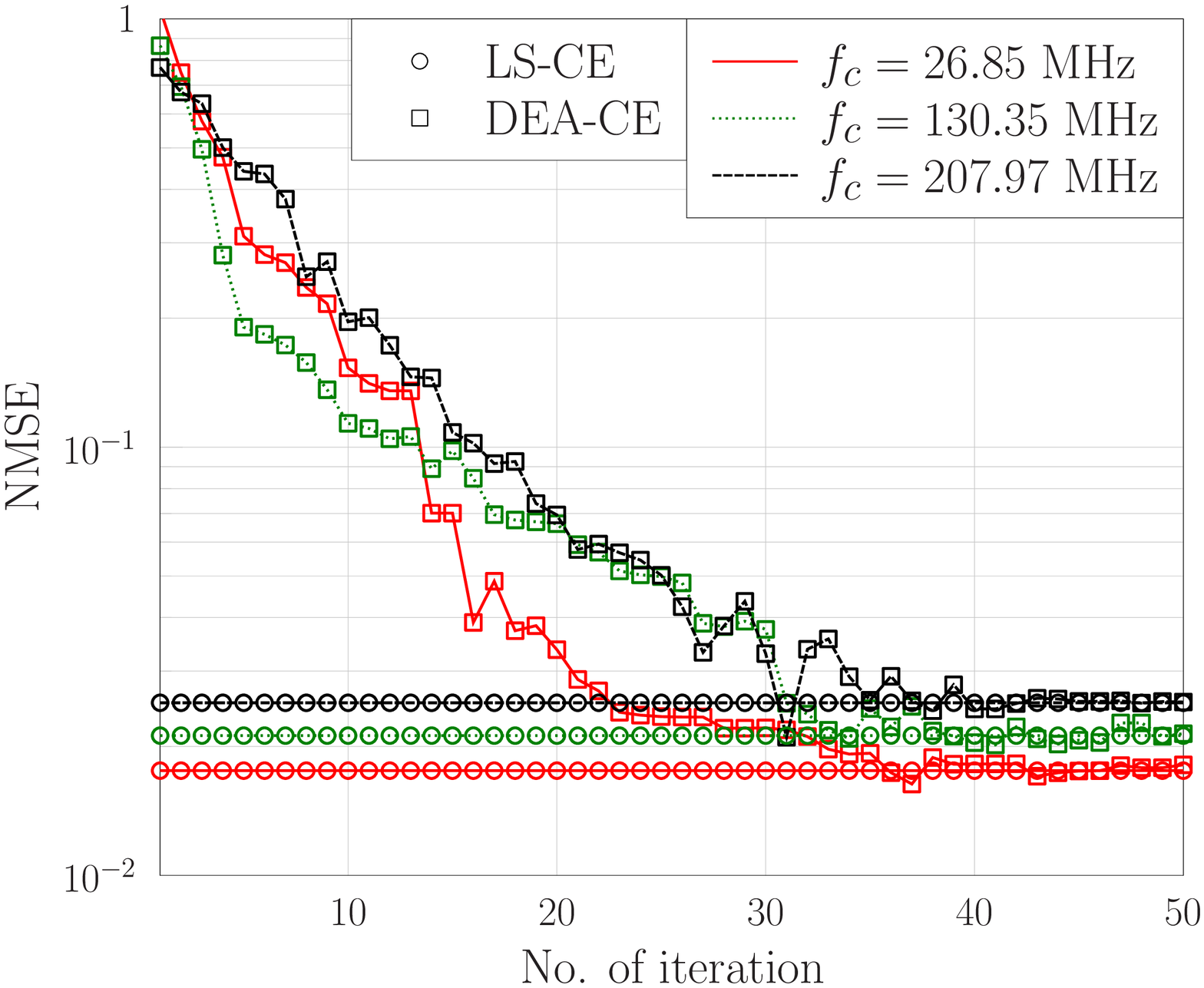}
\end{center}
\vspace{-5mm}
\begin{center}\hspace*{7mm}{\small (a)}\hspace*{80mm}{\small (b)}\end{center}
\vspace{-4mm}
\caption{\small NMSE versus the number of iterations: (a)~$E_b/N_0=20$\,dB, and
 (b)~$E_b/N_0=30$\,dB.}
\label{FIG7}
\vspace{-1mm}
\end{figure*}

\subsection{Per subcarrier NMSE and SER performance of non-turbo CE and MUD}\label{S5.1}

 Let us now consider the non-turbo DEA-CE and DEA-MUD, where  the channel estimator relies
 purely on the pilot symbols only, while the turbo MUD and the decoder perform  iterative
 detection and decoding based on the perfect CSI. This enables us to investigate both the
 NMSE of the DEA-CE and the ideal symbol error ratio (SER) of the DEA-MUD, separately, on
 each subcarrier. Note that the interleaving operation makes it impossible for us to
 investigate the bit error rate (BER) of each individual subcarrier.

 The NMSE performance of the DEA-CE at $E_b/N_0=20$\,dB and 30\,dB is shown in
 Fig.~\ref{FIG5}\,(a) and Fig.~\ref{FIG5}(b), respectively, where $E_b$ is the energy per
 bit and $N_0=\sigma_w^2$ is the noise power. Observe that the DEA-CE achieves an almost 
 identical estimation performance to the LS-CE relying on the same pilot symbols. The idealized
 SER performance based on the perfect CSI achieved by the DEA-MUD at $E_b/N_0=20$\,dB
 and 30\,dB is depicted in Fig.~\ref{FIG6}\,(a) and Fig.~\ref{FIG6}\,(b), respectively,
 in comparison to the idealized SERs of the SUD, ZF-MUD and ML-MUD. As expected, the DEA-MUD
 exhibits a much better detection performance than both the SUD as well as the ZF-MUD, and it is
 capable of approaching the performance of the ML-MUD. Observe that the detection performance
 improvement of the DEA-MUD over both the SUD and over the ZF-MUD is more significant at  high
 frequencies. Interestingly, the SUD outperforms the ZF-MUD at  low frequencies, because
 the latter suffers from serious noise enhancement.

\begin{figure*}[tp!]
\vspace{-1mm}
\begin{center}
\includegraphics[width=0.45\textwidth,angle=0]{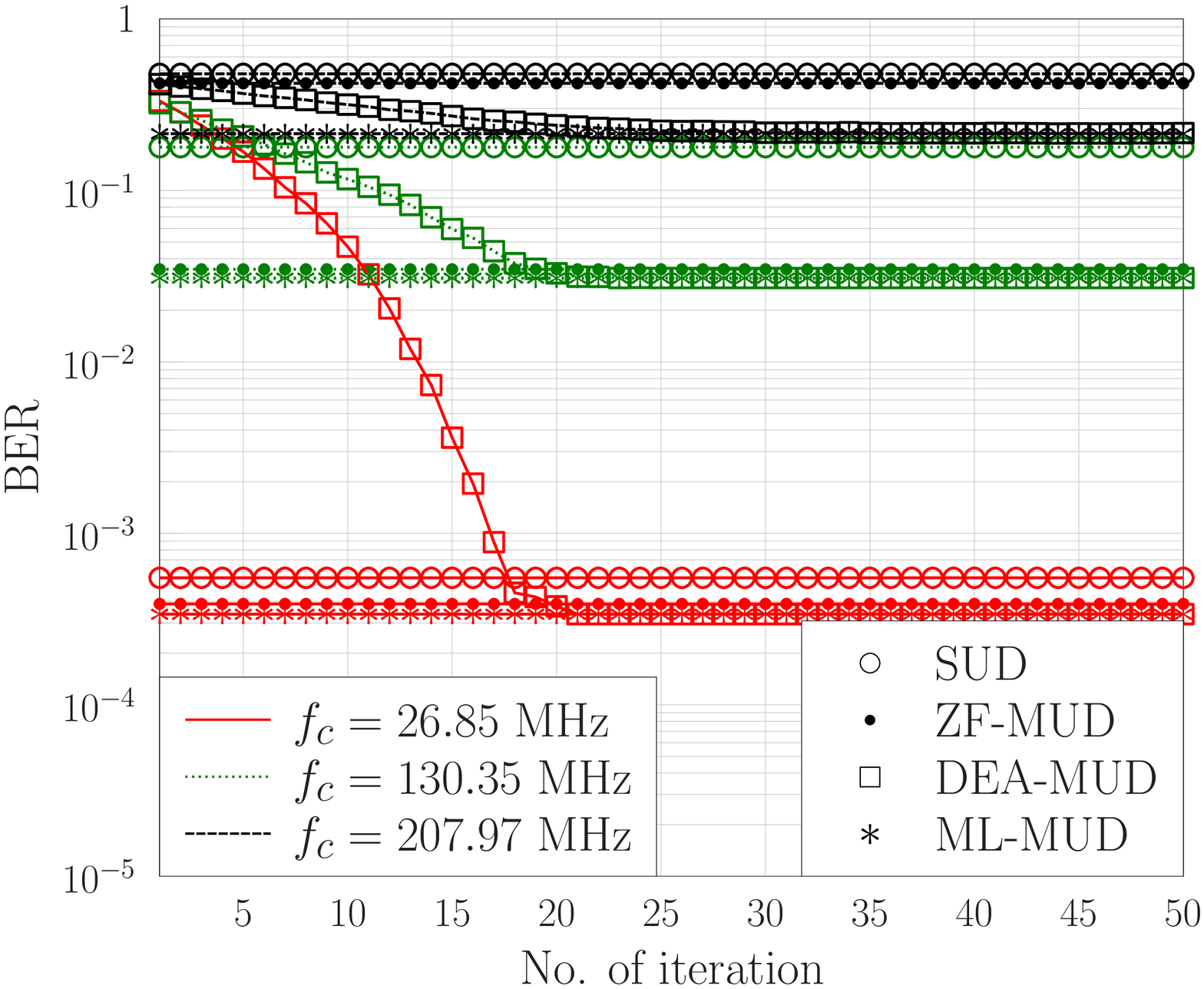}\hspace*{4mm}\includegraphics[width=0.45\textwidth,angle=0]{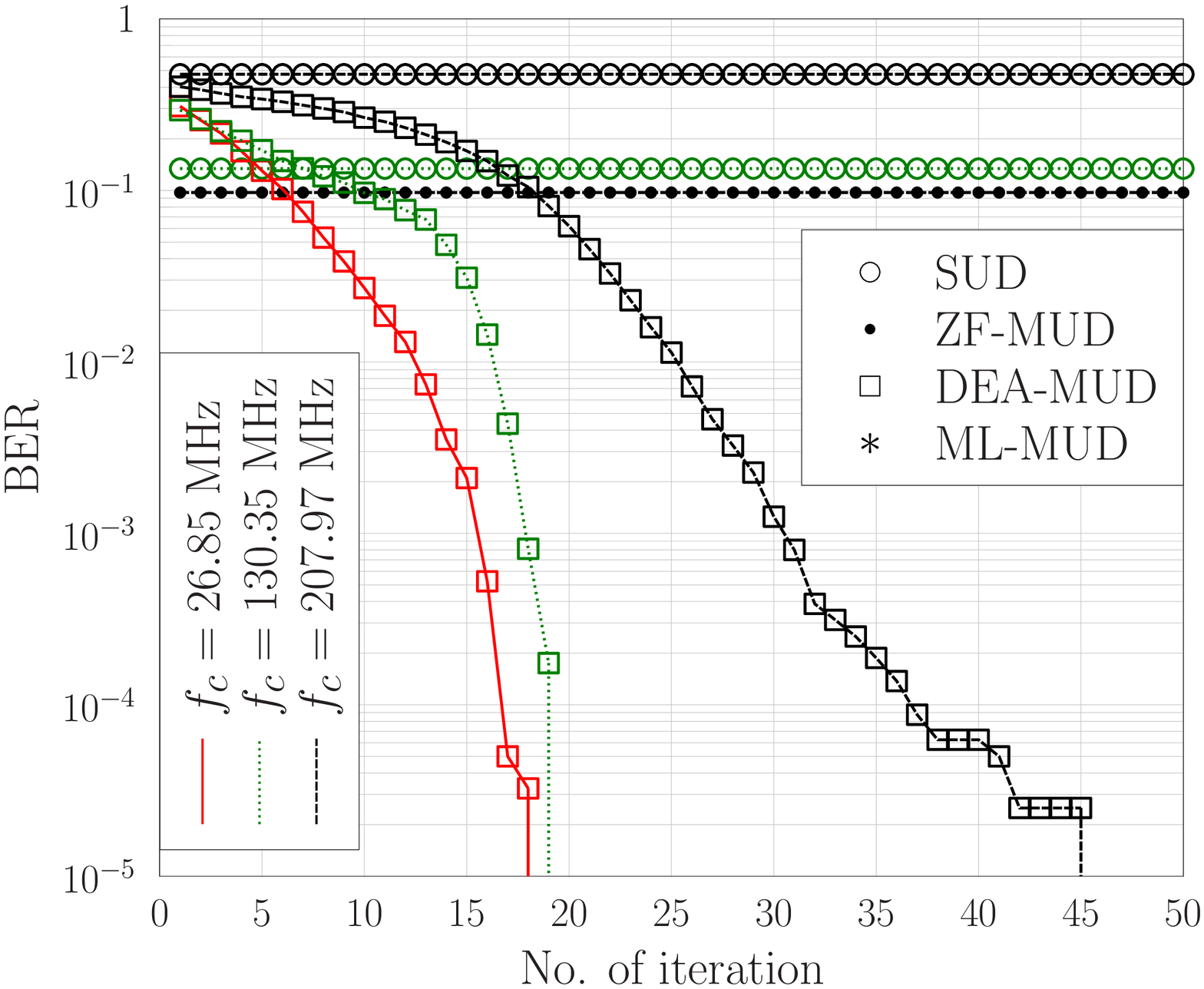}
\end{center}
\vspace{-5mm}
\begin{center}\hspace*{7mm}{\small (a)}\hspace*{80mm}{\small (b)}\end{center}
\vspace{-4mm}
\caption{\small Idealized BER based on perfect CSI versus the number of iterations: (a)~$E_b/N_0=20$\,dB,
 and (b)~$E_b/N_0=30$\,dB.}
\label{FIG8}
\vspace{-4mm}
\end{figure*}

\begin{figure*}[bp!]
\vspace{-3mm}
\begin{center}
\includegraphics[width=0.45\textwidth,angle=0]{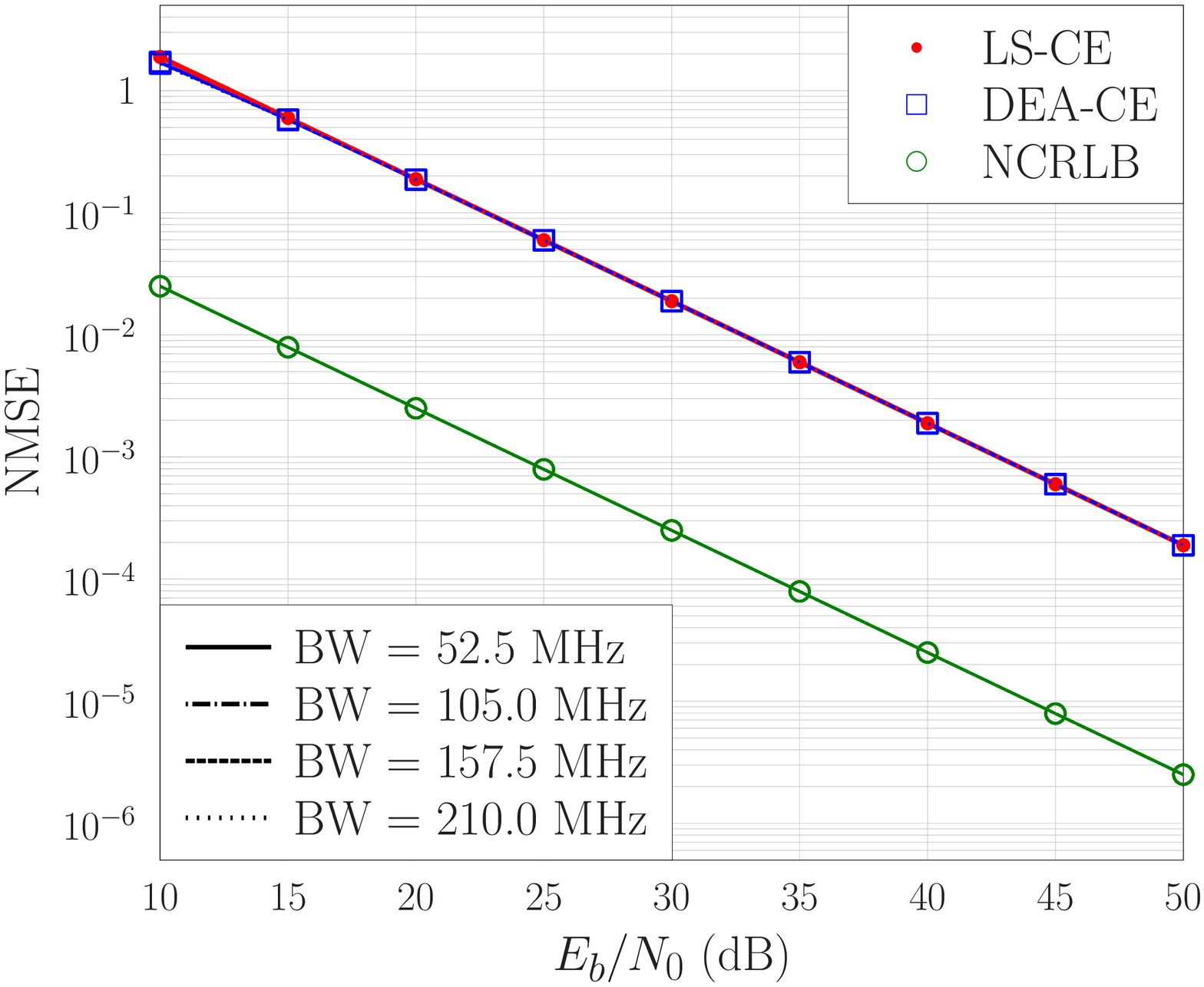}\hspace*{2mm}\includegraphics[width=0.45\textwidth,angle=0]{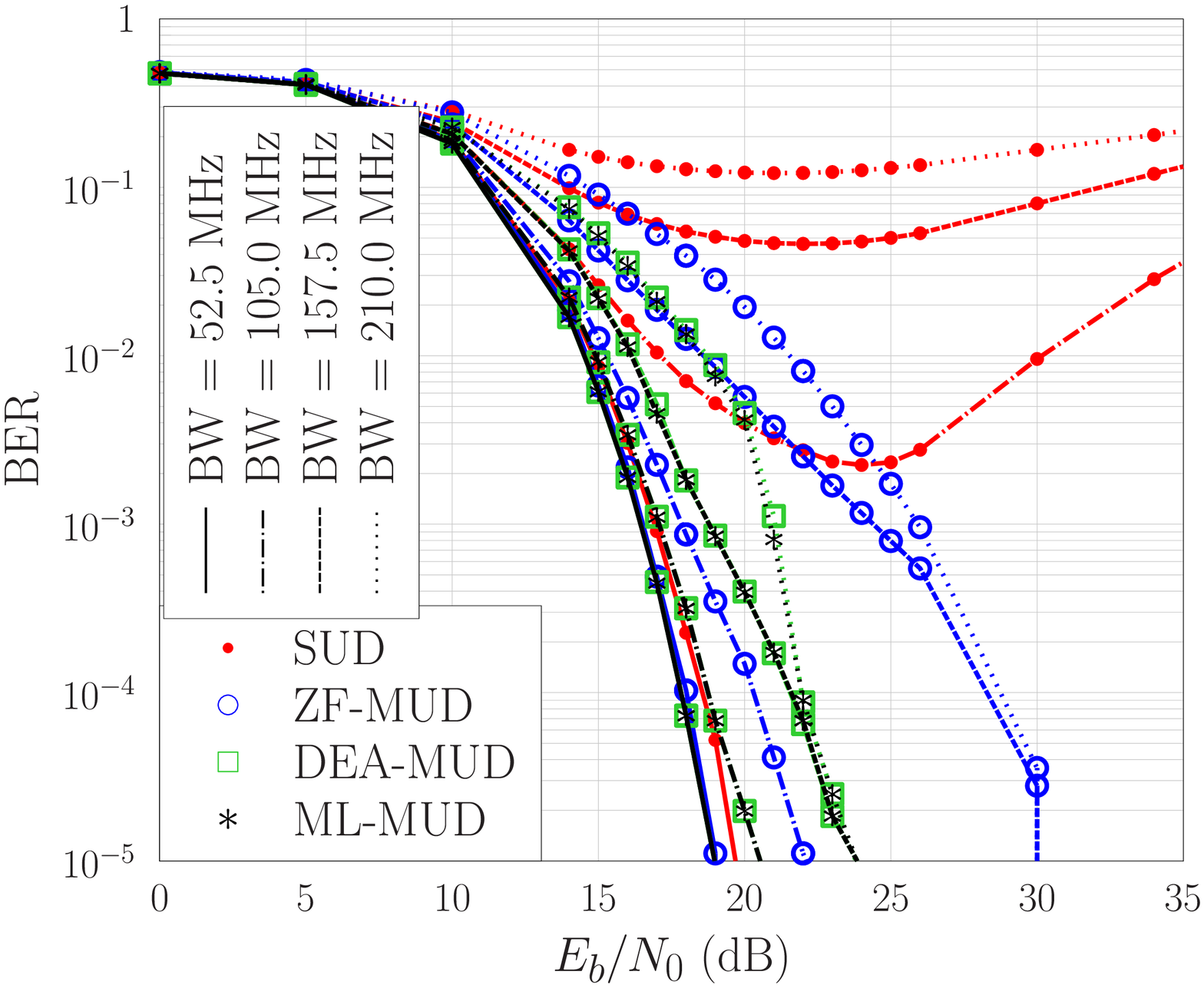}
\end{center}
\vspace{-5mm}
\begin{center}\hspace*{7mm}{\small (a)}\hspace*{80mm}{\small (b)}\end{center}
\vspace{-4mm}
\caption{\small (a)~Training-based NMSE performance versus $E_{b}/ N_{0}$, and (b)~idealized BER
 performance versus $E_{b}/ N_{0}$ relying on perfect CSI, for the system bandwidths of 52.5\,MHz,
 105.0\,MHz, 157.5\,MHz and 210.0\,MHz.}
\label{FIG9}
\vspace{-1mm}
\end{figure*}

\subsection{Convergence of DEA aided CE and DEA aided MUD}\label{S5.2}

 By operating the non-turbo DEA aided CE and DEA aided MUD, we can further investigate
 the convergence of the DEA-CE and the convergence of the DEA-MUD separately. The
 convergence performance of the DEA-CE and the DEA-MUD are shown in Figs.~\ref{FIG7} and \ref{FIG8},
 respectively, at $E_b/N_0=20$\,dB and 30\,dB. Explicitly, we investigate three subcarriers,
 the 500-th, 2500-th and 4000-th tones having  the central frequencies of $f_c=26.85$\,MHz,
 130.35\,MHz and 207.97\,MHz, respectively, where $f_c=26.85$\,MHz is in the frequency
 range of the VDSL2 standard \cite{eriksson2006vdsl2}, while $f_c=130.35$\,MHz and 207.97\,MHz
 are in the frequency range of G.fast.

 It can be seen from Fig~\ref{FIG7} that the DEA-CE converges to the LS-CE solution
 within 30 iterations at $E_b/N_0=20$\,dB and within 40 iterations at  $E_b/N_0=30$\,dB,
 respectively. At $E_b/N_0=20$\,dB, the BERs of the DEA-MUD converge to those of the ML-MUD
 around 20 iterations, as seen from Fig.~\ref{FIG8}\,(a). Note that at $E_b/N_0=30$\,dB, the
 BER of the SUD at $f_c=26.85$\,MHz, the BERs of the ZF-MUD at $f_c=26.85$\,MHz and 130.35\,MHz
 as well as the BERs of the ML-MUD at all the three subcarriers are not included in
 Fig.~\ref{FIG8}\,(b), because they are infinitesimally low based on the perfect CSI. Observe
 from Fig.~\ref{FIG8}\,(b) that at $E_b/N_0=30$\,dB, the DEA-MUD converges to the ML-MUD
 after 18 iterations for $f_c=26.85$\,MHz, 19 iterations for $f_c=130.35$\,MHz and 45
 iterations for $f_c=207.97$\,MHz, respectively.

\begin{figure*}[tp!]
\vspace{-1mm}
\begin{center}
\includegraphics[width=0.49\textwidth,angle=0]{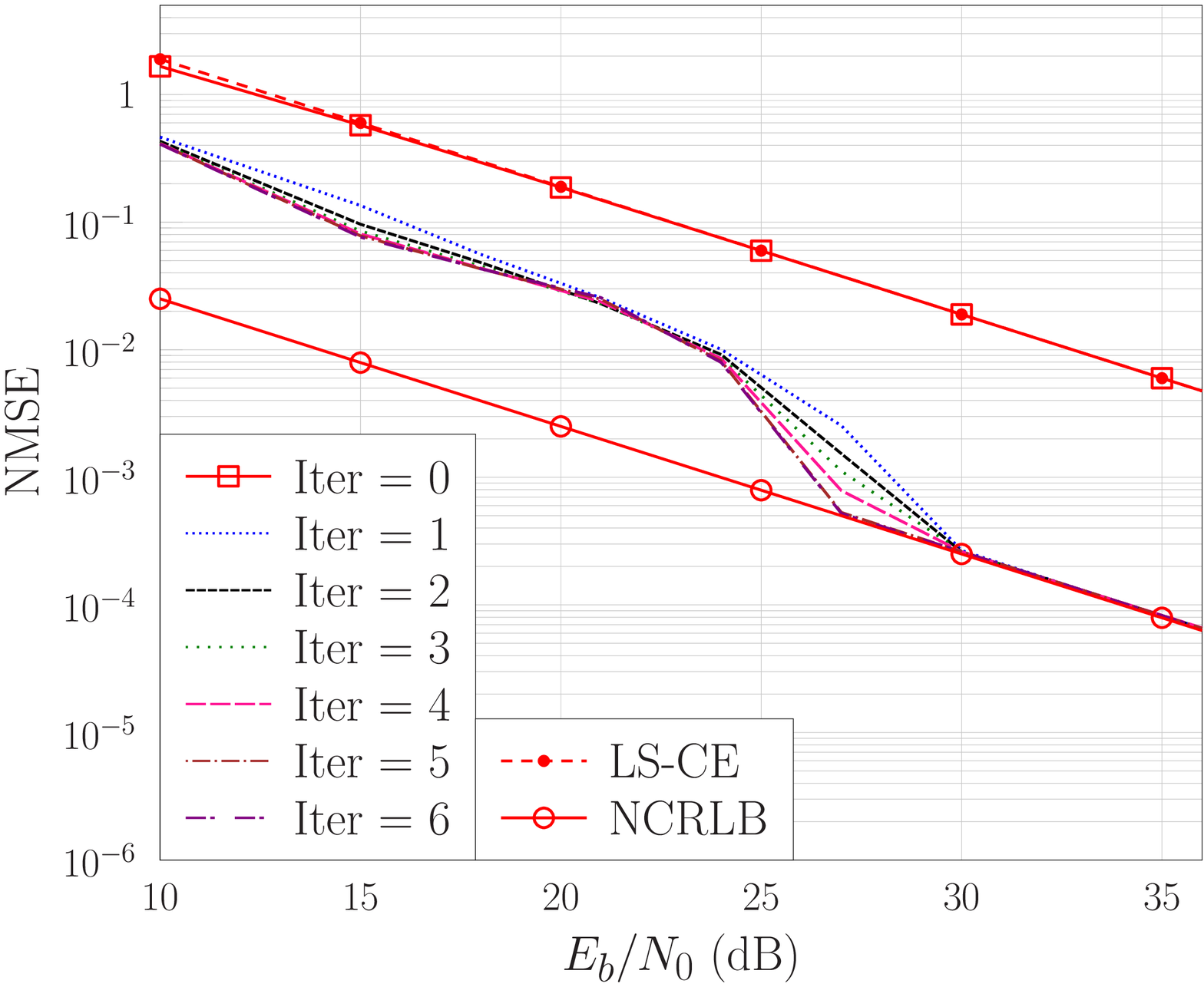}\hspace*{2mm}\includegraphics[width=0.49\textwidth,angle=0]{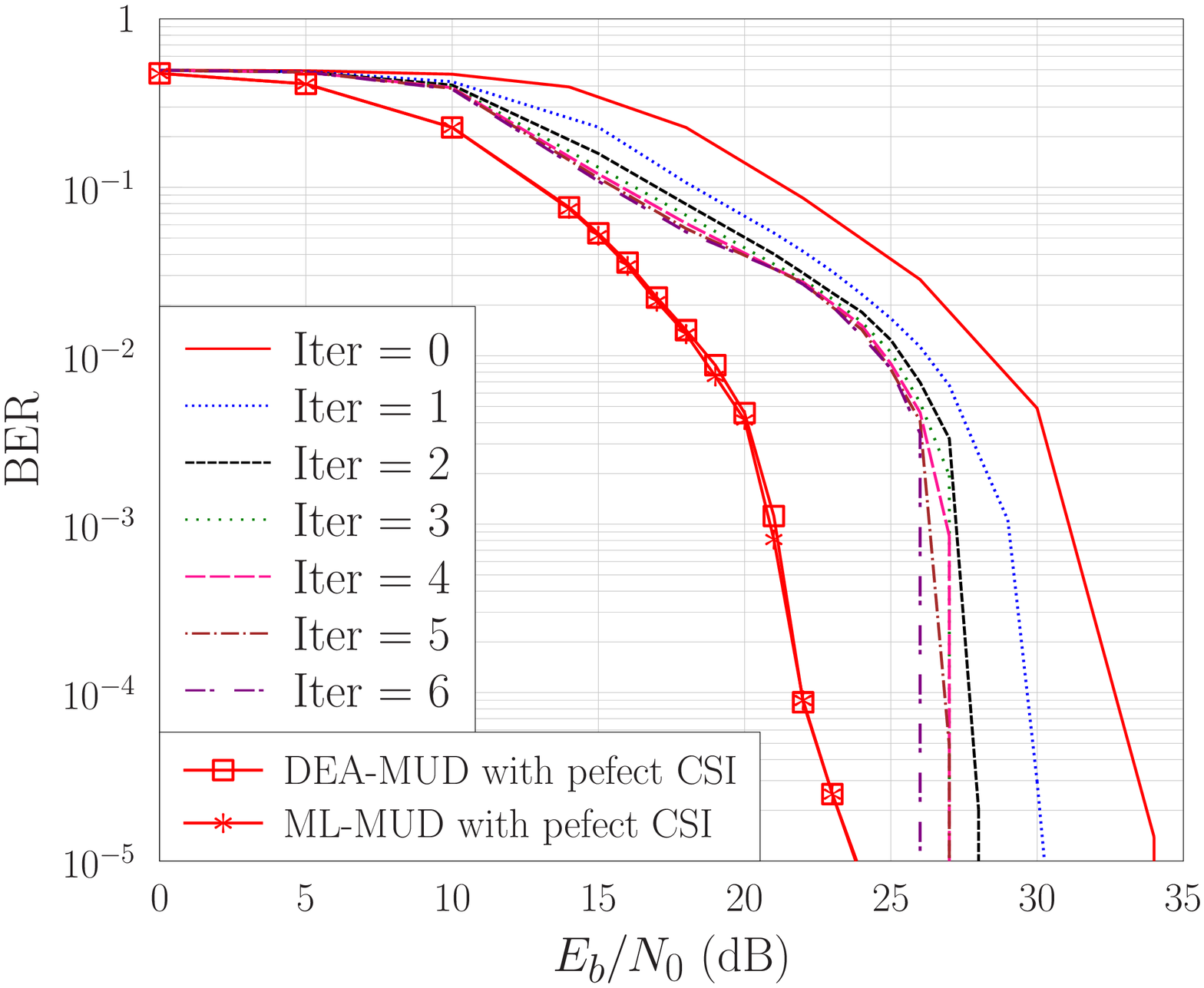}
\end{center}
\vspace{-5mm}
\begin{center}\hspace*{7mm}{\small (a)}\hspace*{80mm}{\small (b)}\end{center}
\vspace{-4mm}
\caption{\small Achievable performance of the DEA aided turbo CE and MUD: (a)~NMSE versus
 $E_{b}/N_{0}$ parametrized by the number of turbo iterations, and (b)~BER versus $E_{b}/N_{0}$
 parametrized by the number of  versus turbo iterations.}
\label{FIG10}
\vspace{-4mm}
\end{figure*}

\subsection{Impact of system bandwidth on the achievable performance}\label{S5.3}

 As mentioned previously, the channel quality is critically dependent on the system's bandwidth
 (BW) or the frequency range. To investigate the influence of the system's frequency range, we
 consider four cases: the 1st case at the BW of 52.5\,MHz covers the first 1024 subcarriers
 of the lowest frequency range, the 2nd case covers the first 2048 subcarriers with the BW of
 105.0\,MHz, the 3rd case has the BW of 157.5\,MHz including the first 3072 subcarriers, and
 the  4th case includes all the 4096 subcarriers and covers the total system BW of 210\,MHz.

 Fig.~\ref{FIG9}\,(a) depicts the NMSE performance of the DEA-CE based on the  pilot symbols.
 As expected, the NMSE of the DEA-CE is identical to that of the LS-CE, and the system BW has
 no impact on the training performance of an unbiased channel estimator. Observe from
 Fig.~\ref{FIG9}\,(a) that there is an approximately 18\,dB gap between the training-based NMSE
 and the NCRLB, where the NCRLB is calculated based on both the pilots and the data.
 Fig.~\ref{FIG9}\,(b) shows the idealized BER performance of the DEA-MUD based on the perfect
 CSI for different system bandwidths. As expected, the BER performance is better for the system
 having a lower frequency range, because the channel quality at a lower frequency range is
 better. The results of Fig.~\ref{FIG9}\,(b) also confirm that the DEA-MUD and the ML-MUD
 exhibit almost identical detection performance, and the DEA-MUD outperforms both the SUD and
 the ZF-MUD, particularly for the systems including higher frequencies. For example, for the system
 including all the 4096 subcarriers, the DEA-MUD attains  an approximately 7\,dB SNR gain over
 the ZF-MUD at the BER level of $10^{-5}$. For this system, the SUD exhibits a high error floor.

\subsection{Achievable performance of the DEA aided turbo CE and MUD}\label{S5.4}

 We now investigate the achievable performance of the proposed DEA aided turbo CE and MUD.
 Again, we consider the system relying on all the 4096 subcarriers and having the total system
 bandwidth of 210\,MHz. The NMSE and BER versus $E_{b}/N_{0}$ performance of this DE aided
 turbo CE and MUD scheme is parametrized by the number of turbo iterations, as depicted in
 Fig.~\ref{FIG10}\,(a) and Fig.~\ref{FIG10}\,(b), respectively.

 The results of Fig.~\ref{FIG10} need further explanations. Initially, given the training
 pilots, the DEA-CE estimates the channel, and the NMSE of this channel estimate is given
 by the $\text{Iter}=0$ curve of Fig.~\ref{FIG10}\,(a), which is identical to that of the
 training based LS-CE. Given this estimate, the DEA-MUD and the channel decoder perform
 detection and decoding by iteratively exchanging soft extrinsic information, and after
 convergence, the detected data exhibits the BER represented by the $\text{Iter}=0$ curve
 of Fig.~\ref{FIG10}\,(b). The detected data are then fed back to the DEA-CE, which
 carries out the next CE iteration based on both the detected data and pilots, leading to
 the improved NMSE as seen from Fig.~\ref{FIG10}\,(a). The enhanced estimated CSI is in
 turn exploited by the DEA-MUD/channel decoder for producing  the detected data at an even
 lower BER. This `turbo' procedure continues until at the 6th iteration and around the SNR
 of 27 dB, the BER of  the detected data becomes infinitesimally low. Observe from
 Fig.~\ref{FIG10}\,(a) that at this point, the NMSE of the DEA aided turbo CE and MUD
 approaches the NCRLB. {
Explicitly, at the 6th iteration and around the SNR of
27 dB, the achievable NMSE has approached the NCRLB. 
}
This is because at this point, the detected data becomes the true
 data. Not surprisingly, at this point, the BER of the DEA aided turbo CE and MUD approaches
 that of the idealized ML-MUD associated with  perfect CSI, which is also identical to that
 of the idealized DEA-MUD relying on perfect CSI.

 It is worth emphasizing the significance of the iterative gain obtained. Specifically, by
 iteratively exchanging extrinsic information between the continuous DEA aided CE and the
 discrete DEA aided MUD, approximately 18\,dB of NMSE gain as attained for the channel
 estimator and around 10\,dB  of SNR gain is achieved for the MUD.

\begin{figure}[tp!]
\vspace{-1mm}
\begin{center}
 \includegraphics[width=0.48\textwidth,angle=0]{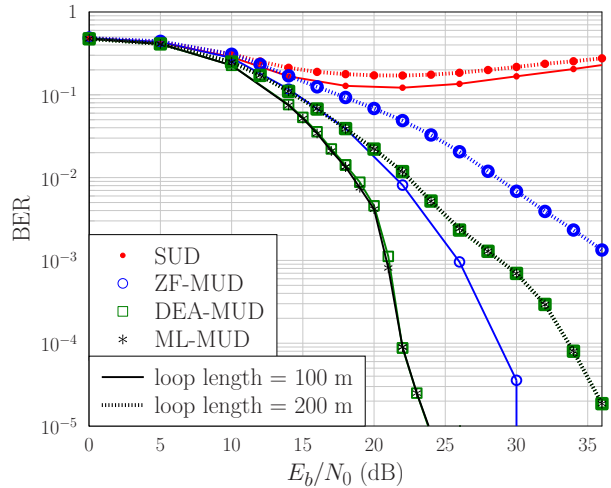}
\end{center}
\vspace{-4mm}
\caption{\small Impact of loop length on the achievable detection performance of four
 idealized MUDs associated with perfect CSI.}
\label{FIG11}
\vspace{-4mm}
\end{figure}

\subsection{Impact of loop length}\label{S5.5}

 Although the lengths of the users' DSL lines are different in practical deployments,
 for the convenience of investigating the impact of loop length, we assume that all
 the users have the same loop length. Fig.~\ref{FIG11} shows the influence of the
 loop length of DSL lines on  the achievable BER performance of the four idealized
 MUDs based on perfect CSI. It can be clearly seen from Fig.~\ref{FIG11} that
 increasing the loop length significantly degrades the achievable BER performance.
 Observe furthermore from Fig.~\ref{FIG11} that the DEA-MUD attains the optimal
 performance of the ML-MUD, and it considerably outperforms the ZF-MUD. For these two
 systems, the SUD exhibits high BER floors.

\subsection{Impact of impulse noise}\label{S5.6}

 Next we investigate the impact of impulse noise on the achievable BER performance.
 When the additive impulse noise is taken into account, the system model (\ref{EQ1_re_sig})
 can be rewritten as
\begin{align}\label{EQ38_re_sig_im} % eq32
 \boldsymbol{Y} =& \boldsymbol{H} \boldsymbol{X} + \boldsymbol{W} + \boldsymbol{U} ,
\end{align}
 where $\boldsymbol{U}\in \mathbb{C}^L$ is the impulse noise vector. An OFDM symbol is
 infected by the impulse noise with the probability of $\kappa$, where typically we have 
 $\kappa\in [0.01, ~ 0.1]$ \cite{kirkby2001text,ndo2013markov}. Note that an OFDM symbol
 includes the data at all the tones, $\boldsymbol{X}[1],\cdots ,\boldsymbol{X}[4096]$,
 where we have re-introduced the omitted tone index of $\boldsymbol{X}$. We assume that
 the impulse noise $\boldsymbol{U}$ obeys the complex Gaussian distribution associated 
 with a zero-mean vector and the covariance matrix $\sigma^2_u\boldsymbol{I}_L$.
 Typically, the impulse noise is 20\,dB stronger than the additive Gaussian noise
 $\boldsymbol{W}$ \cite{neckebroek2013comparison,bai2017discrete}, that is, $10\log_{10}
 \left(\sigma_u^2/\sigma_w^2\right)=20$\,dB. In our simulations, we consider both
 $\kappa =0.01$ and $\kappa =0.1$, which can be viewed as the lower bound and the upper
 bound of the probability that an OFDM symbol is infested by impulse noise.

\begin{figure}[tp!]
\vspace{-1mm}
\begin{center}
 \includegraphics[width=0.48\textwidth,angle=0]{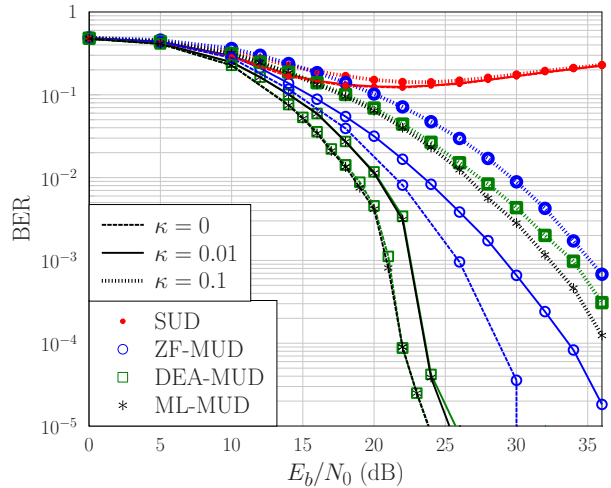}
\end{center}
\vspace{-4mm}
\caption{\small Impact of impulse noise on the achievable detection performance
 of four idealized MUDs associated with perfect CSI.}
\label{FIG12}
\vspace{-1mm}
\end{figure}

 It can be seen from Fig.~\ref{FIG12} that the impulse noise degrades the achievable
 detection performance, particularly for high  $\kappa$, where the case of $\kappa =0$
 corresponds to no impulse noise. It can also be seen that at $\kappa =0.01$, both the
 ML-MUD and the DEA-MUD exhibit almost an identical performance, while at $\kappa =0.1$,
 the DEA-MUD is slightly inferior to the ML-MUD. Not surprisingly, the DEA-MUD considerably
 outperforms the ZF-MUD.

\begin{figure}[tp!]
\vspace{-1mm}
\begin{center}
 \includegraphics[width=0.48\textwidth,angle=0]{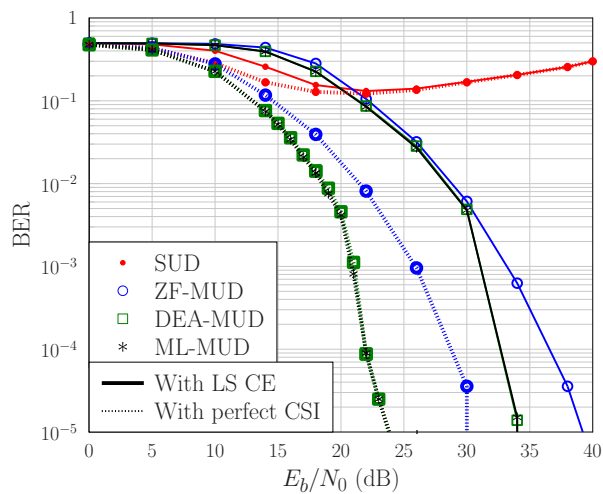}
\end{center}
\vspace{-4mm}
\caption{\small Comparison of the achievable detection performance of four MUDs based on
 LS-CE acquired by training as well as based on perfect CSI.}
\label{FIG13}
\vspace{-6mm}
\end{figure}

\subsection{Impact of channel estimation error}\label{S5.7}

 In Fig.~\ref{FIG13}, we compare the BER performance of the four MUDs based on the LS-CE
 acquired by training to those associated with perfect CSI. Clearly, the channel estimation
 error has a significant impact on the achievable detection performance of an MUD.
 Explicitly, for both the DEA-MUD and the ML-MUD, there exists an SNR gap of around 10\,dB
 between the idealized performance associated with the perfect CSI and the performance
 associated with the estimated CSI, which can also be seen from Fig.~\ref{FIG10}\,(b). It
 can also be seen from Fig.~\ref{FIG13},  that the SNR gap is approximately 9\,dB between
 the idealized ZF-MUD associated with   perfect CSI and the ZF-MUD based  estimated CSI. 
 
\subsection{Computational complexity comparison}\label{S5.8}

 As demonstrated by the aforementioned results, given the same CSI, the DEA-MUD is
 capable of attaining the optimal detection performance of the ML-MUD. We now compare 
 the computational complexity of the DEA-MUD to that of the ML-MUD. Again, the ML-MUD finds
 the optimal solution by evaluating the CF values of all the $M^L=16^4=65536$ potential
 candidates on each single tone. By contrast, the discrete DEA evolves a population of
 `candidates', as detailed in Section~\ref{S3.3}, based on the CF values of the population.
 Let $N_{\rm ML}$ be the total number of CF evaluations imposed by the ML-MUD  and let
 $N_{\rm DEA}$ be the total number of CF evaluations required by the DEA to converge to the
 optimal ML solution. We can compare the computational complexities of both  the DEA-MUD 
and of the ML-MUD by calculating the ratio
\begin{align}
 \text{Complexity of DEA-MUD} =& \frac{N_{\rm DEA}}{N_{\rm ML}} ~ [\%] ,
\end{align}
 which we use to quantify the computational complexity of the DEA-MUD, in comparison to
 the ML-MUD.

\begin{figure}[tp!]
\vspace{-1mm}
\begin{center}
 \includegraphics[width=0.48\textwidth,angle=0]{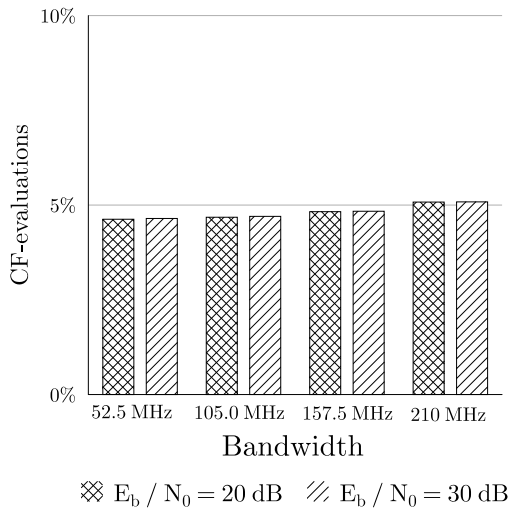}
\end{center}
\vspace{-8mm}
\caption{\small Computational complexity of the DEA-MUD expressed as the ratio of its
 required CF evaluations over the total CF evaluations imposed by the ML-MUD. The
 perfect CSI is assumed.}
\label{FIG14}
\vspace{-4mm}
\end{figure}

 Fig.~\ref{FIG14} compares the complexities of the DEA-MUD to those of the ML-MUD for
 the same four systems as specified in Section~\ref{S4.3} at both $E_b/N_0= 20$\,dB and
 30\,dB. It can be seen from Fig.~\ref{FIG14} that the DEA-MUD only requires 5\% of the
 computational complexity imposed by the ML-MUD, while still attaining the same optimal
 solution as the ML-MUD.

\section{Conclusions}\label{S6}

 We have proposed a DEA aided turbo CE and MUD for mitigating the adverse effects of
 FEXT encountered by the G.Fast systems caused by the utilization of high  frequencies up
 to 212\,MHz. The proposed DEA aided turbo CE and MUD is constituted by a continuous DEA
 aided CE and a discrete DEA aided MUD, exchanging extrinsic information between them. We
 have demonstrated that our DEA-MUD significantly outperforms the  widely adopted
 low-complexity ZF-MUD, also known as the ZF-FEXT canceller. More remarkably, we have shown
 that given the same CSI, our discrete DEA aided MUD is capable of attaining the optimal
 performance of the ML-MUD, while only imposing 5\% of the computational complexity
 associated with the ML-MUD. In our simulation study, we have also investigated the impact 
 of CE error, of the impulse noise and of the loop length. Most importantly, in this paper,
 we have demonstrated that by iteratively exchanging information between the continuous
 DEA aided CE and the discrete DEA aided MUD, the DEA-CE is capable of approaching the
 optimal CRLB of the channel estimate, while the DEA-MUD based on the estimated CSI is
 capable of attaining the optimal detection performance of the idealized ML-MUD associated
 with   perfect CSI. Specifically, we have shown that 18\,dB of the NMSE gain is attained
 by the channel estimator and 10\,dB of the SNR gain is gleaned by the MUD by exploiting
 iteration gains. This study therefore has demonstrated that the proposed  DEA aided
 turbo CE and MUD  is capable of offering near-capacity performance at an affordable
 complexity for the emerging G.fast systems. 

\section*{Acknowledgements}
The authors would like to thank Tong Bai from Next Generation Wireless group for his discussions and contributions about impulse noise.

\bibliographystyle{IEEEtran}
%\small \bibliography{Myreference_Article_2016_2_5}

\end{document}